\documentclass[final]{article}

\usepackage{arxiv}

\usepackage[utf8]{inputenc} 
\usepackage[T1]{fontenc}    
\usepackage{hyperref}       
\usepackage{url}            
\usepackage{booktabs}       
\usepackage{amsfonts}       
\usepackage{nicefrac}       
\usepackage{microtype}      
\usepackage{lipsum}
\usepackage{graphicx}

\usepackage{xcolor}
\usepackage{tikz}
\usepackage{amsmath}
\usepackage{amssymb}

\usepackage{lineno}
\usepackage{algorithm2e}
\usepackage{textcomp} 
\usepackage{subcaption}
\usepackage{adjustbox}
\usepackage{url}

\modulolinenumbers[5]
\usetikzlibrary{calc}
\usetikzlibrary{intersections}
\usetikzlibrary{through}

\title{Geospatial Perspective Reprojections for Ground-Based Sky Imaging System}

\author{
 Guillermo Terr\'en-Serrano \\
  Department of Electrical and Computer Engineering \\
  The University of New Mexico \\
  Albuquerque, NM 87131, United States\\
  \texttt{guillermoterren@unm.edu} \\
 \And
  Manel Mart\'inez-Ram\'on \\
  Department of Electrical and Computer Engineering \\
  The University of New Mexico \\
  Albuquerque, NM 87131, United States\\
  \texttt{manel@unm.edu} \\
}

\begin{document}

\maketitle

\begin{abstract}
    Sky imaging systems use lenses to acquire images concentrating light beams in a sensor. The light beams received by the sky imager have an elevation angle with respect to the device normal. Thus, the pixels in the image contain information from different areas of the sky within the imaging system field of view. The area of the field of view contained in the pixels increases as the elevation angle of the incident light beams decreases. When the sky imager is mounted on a solar tracker, the light beam's angle of incidence in a pixel varies over time. This investigation formulates and compares two geospatial reprojections that transform the original euclidean frame of the imager plane to the geospatial atmosphere cross-section where the sky imager field of view intersects the cloud layer. One assumes that an object (i.e., cloud) moving in the troposphere is sufficiently far so the Earth's surface is approximated \emph{flat}. The other transformation takes into account the curvature of the Earth in the portion of the atmosphere (i.e., voxel) that is recorded. The results show that the differences between the dimensions calculated by both geospatial transformations are in the order of magnitude of kilometers when the Sun's elevation angle is below $30^\circ$.
\end{abstract}
    
\keywords{Infrared Camera \and Perspective Reprojection \and Sky Imaging \and Solar Forecasting; Sun Tracking}

\section{Introduction}

The Global Solar Irradiance (GSI) that reaches the Earth's surface depends on shadows projected by moving clouds in the troposphere \cite{TZOUMANIKAS2016}. Consequently, clouds influence the energy generation in Photovoltaic (PV) powered smart grids. GSI forecasting methods, which are efficient for intra-hour horizons, analyze the dynamics of clouds to predict GSI minutes ahead of time using data acquired using ground-based sky imagers \cite{KONG2020}, to control the storage and dispatch of energy.

The horizons of  intra-hour solar forecasting depend on the Field of View (FOV) of the sky imager used to acquire the images. A sky imager may be composed of one or multiple visible or Infrared (IR) imagers, or both, and their FOV generally varies from $60^\circ$ (low) to $180^\circ$ (large). However, unless the sky imager is mounted on a solar tracker \cite{MAMMOLI2013, CHU2016, TERREN2020c}, the necessary FOV to perform an accurate intra-hour solar forecast is large. Total Sky Imagers (TSI) achieved large FOV sky images using a concave mirror to reflect light beams into a visible \cite{Chow2011} or IR camera \cite{Redman2018}, and the camera is installed on a support at the focal distance of the mirror \cite{Gohari2014, Marquez2013}. An alternative to reflective sky imagers (in visible light sky images), is to increase the camera's FOV using a fisheye lens \cite{Li2012, Fu2013, Liu2015, Cheng2017a}. These are generally known as ``all sky imagers’’ \cite{Shi2019, CALDAS2019, HASENBALG2020}. Similarly, the FOV of IR sky imagers can be enlarged applying image processing techniques to merge images acquired from multiple low FOV imagers \cite{MAMMOLI2019}.

Each of these sky imagers use light beams received at an angle with respect to the imager's plane. Therefore, the produced distortion should be corrected using a geometric transformation to compute the velocity vectors of a cloud. The geometric transformation proposed by \cite{Nummikoski2013} transforms the Euclidean coordinate system of the pixels to a coordinate system based on the azimuth and elevation angles. This transformation was implemented by \cite{richardson2017} for reprojecting the pixels of a TSI, in the atmosphere cross-section plane, using height measurements acquired using a nearby ceilometer. Ceilometers estimate the height of clouds and have been used to validate low-cost approaches to approximate the height of a cloud using multiple all sky imagers \cite{nguyen2014, KUHN2018}. However, this device is expensive and it is not applicable to more general operations such as a cloud speed sensor \cite{wang2016}. Another low-cost alternative to determine the velocity of clouds moving in the atmosphere cross-section, and thus estimating their heights, was developed using an all sky imager and a grid of sensors (i.e., pyranometers) by \cite{wang2019}.

Nevertheless, these geometric transformations were developed for static sky imagers (i.e., TSI and all sky imager). In contrast, the geospatial reprojections introduced in this investigation not only work for static sky imagers, but are also applicable to sky imagers mounted on a solar tracker. In this last case, the perspective in the images is a function of the Sun's elevation and azimuth angles. The first approximation is a reprojection for devices that do not record low elevation angles (see section~\ref{sec:flat_earth}), while the second computes accurate reprojections even when the elevation angle is low (see section~\ref{sec:great_circle}). The proposed reprojections were originally developed for a low FOV sky imager mounted on a solar tracker \cite{TERREN2020d, TERREN2020b}, however, it is possible to obtain the geospatial reprojection for any FOV and elevation angle by reparameterizing the algorithms. As a ceilometer was not available, the proposed methods were developed so that ceilometer measurements are not required.

\section{Rectilinear Lens}

The  acquired image is the light beam  refraction in a converging point of the emitted black body radiation. The image resolution is defined as $N \times M$ pixels. If the radiant objects (the Sun and the clouds) are at a distance $z \rightarrow \infty$, the radiation rays converge at the focal length. Consequently, 
\begin{equation}
    \frac{1}{f} = \frac{1}{z} + \frac{1}{D}\approx \frac{1}{D},
\end{equation}
where $f$ is the focal length and $D$ is the distance from the lens to the converging point. 
The relation between the diagonal FOV and the focal length $f$ for a rectilinear lens is
\begin{equation}
    \tan \frac{\mathrm{FOV}}{2} = \delta \frac{N_{diag}}{2f},
\end{equation}
where $N_{diag} = \sqrt{N^2 + M^2}$ is the number of pixels in the diagonal of the sensor an $\delta$ is the pixel size. Therefore, the focal length $f$ of camera is,
\begin{equation}
    f = \frac{\delta}{2} \frac{N_{diag}}{\tan \frac{\mathrm{FOV}}{2}}.
\end{equation}

\section{Flat Earth Approximation}\label{sec:flat_earth}

\begin{figure}[!htb]
    \centering
    \resizebox{12.5cm}{!}{\input{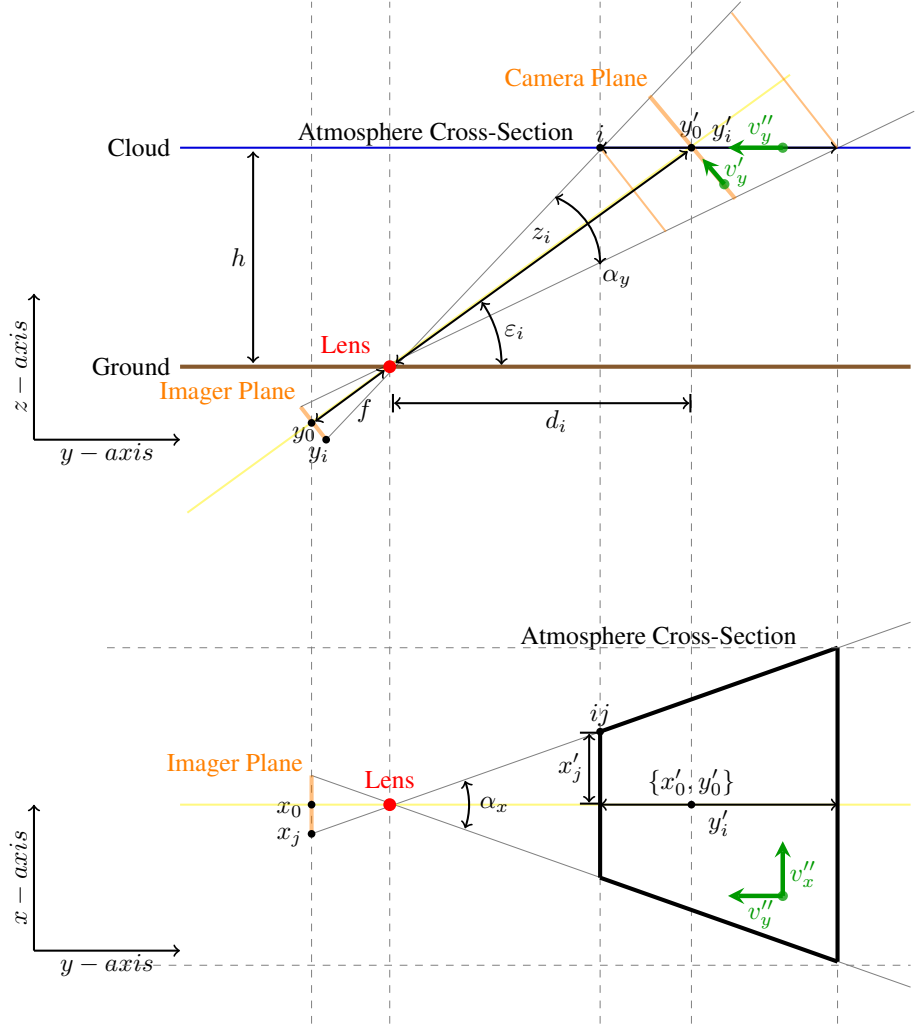}}
    \caption{\emph{Flat} Earth approximation of the geospatial reprojection. Top pane (side view): the reprojection depends on the distance $z_i$ of an object to the imager, the height $h$ and the elevation $\varepsilon_i$. Bottom pane (top view): relation of the angular increments $\alpha$ used to compute the elevation angle $\varepsilon_i$ of each one of the pixels $i,j$ in the image. The velocity decomposition $\mathbf{v}^{\prime} = \{v^{\prime}_{x}, v_{y}^{\prime}\}$ shows that cloud velocity components have a perspective distortion in the x-axis and in the y-axis, due to the camera plane inclination of $\varepsilon$ degrees with respect to the normal. $x_j^\prime$ and $y_i^\prime$ represent the coordinates of the pixel in the image (see in Eq.~6). When the coordinate system is centered applying Eq.~\ref{eq:origen}, $\mathbf{x}_0 = \{x_0^\prime, y_0^\prime$\} represent the origin of coordinates.}
    \label{fig:flat_earth}
\end{figure}

The \emph{flat} Earth approximation is viable without large error (when the elevation of the Sun $\varepsilon_0$ is higher than $30^\circ$) because the portion of the Earth's atmosphere in the FOV of the camera is much smaller than its entire surface. With this assumption, the reprojection from the sensor plane to the atmosphere cross-section plane (in Fig.~\ref{fig:flat_earth}) is obtained with the distance $z$ of a cloud to the camera lens. The distance $z$ is a function of the cloud height $h$ and the elevation angle $\varepsilon$ of the cloud in a pixel,
\begin{equation}\label{eq:distance_z}
    z = \frac{h}{\sin \varepsilon}.
\end{equation}

The reprojection is computed with respect to the coordinates of each pixel $i,j$ in the imager plane. The coordinates of a pixel in the imager plane are defined as $x_{j} = j \delta$ and $y_{i} = i \delta$. In this reprojection, we assume that the elevation angle $\varepsilon_i$ is different in each row  $i$ and constant in each column $j$ of pixels in an image, and the differential angle $\alpha_j$ (formed by the position of Sun and a pixel) is different in each column $j$ and constant in each row $i$ of pixels. This assumption is valid since the FOV of the individual pixels in the rectilinear lens is sufficiently small. As seen in Fig.~\ref{fig:flat_earth}, when intersecting a cloud layer, the projection of the 3D pyramid defined by the camera FOV in a 2D plane forms a triangle. The elevation $\varepsilon_i$ and azimuth $\alpha_j$ angles for each pixel $i,j$ are,
\begin{equation}
    \begin{split}
        \boldsymbol{\varepsilon} &= \left\{ \left( \varepsilon_0 + i \frac{\nu}{2} \right) \ \middle| \ \varepsilon_i \in \mathbb{R}^{(0, \pi]}, \ \forall i = - \frac{N}{2}, \ \ldots,\ \frac{N}{2} \right\},\\
        \boldsymbol{\alpha} &= \left\{ \left( \alpha_0 + j \frac{\nu}{2}\right) \ \middle| \ \alpha_j \in \mathbb{R}^{(0, \alpha_x/2]}, \ \forall j = - \frac{M}{2}, \ \ldots,\ \frac{M}{2} \right\},
    \end{split}  
\end{equation}
where $\nu = [\mathrm{FOV}/\sqrt{N^2 + M^2}] \cdot [ \pi/180 ]$ is the camera ratio in radians per pixel, $\varepsilon_0$ is the Sun's elevation angle, and $\alpha_0 = 0$. Therefore, $\alpha_j = 0$ and $\varepsilon_i = \varepsilon_0$ represent the center of the image (since $\alpha_0 = 0$), but only when the number of pixels $N$ and $M$ are odd numbers. For all pixels, $\nu$ is approximated by a constant. In this way, the FOV is $\alpha_x = \nu M $ and $\alpha_y = \nu N$ in the $x$ and $y$ axis respectively.

The length of a row of pixels $j$ reprojected in the atmosphere cross-section is $x^{\prime}_{i,j} = x_j \cdot z_i /f $, so substituting $z_i$ in Eq.~\ref{eq:distance_z}, the coordinates of the imager plane reprojected in the atmosphere cross-section are,
\begin{equation}\label{eq:flat_earth_coordinates}
    \begin{split}
        x^\prime_{i,j} &= \frac{x_j}{f} \cdot z_i = \frac{x_j}{f} \cdot \frac{h}{ \sin \varepsilon_i} \\
        y^\prime_{i}   &= \frac{y_i}{f} \cdot z_i = \frac{y_i}{f} \cdot \frac{h}{\sin \varepsilon_i}.
    \end{split}
\end{equation}

\section{Great Circle Approach}\label{sec:great_circle}

The atmosphere cross-section plane can be approximated more exactly using the pyramid formed by the camera FOV when intersects a cloud layer at height $h$ in point $D$ in Fig.~\ref{fig:great_circle}. The assumption is that the Earth and the cloud layer surface are two perfect spheres. The \emph{great} circle is defined as the cloud layer surface at height $h$, and \emph{small} circle is the Earth's surface. The tangent plane to the Earth's surface which intersects with the cloud layer is the chord $AB$ (see Fig.~\ref{fig:great_circle}). The Earth’s radius is $r_{Earth}$. The sagitta $\ell_i = h - v_i$ is the length from the middle of chord $C_iD_i$ to the cloud layer, and $v_i$ is the perpendicular distance from the \emph{great} circle to the \emph{small} circle. The \emph{great} and \emph{small} circles radii are respectively,
\begin{equation}
    \begin{split}
        R &= r + h \\
        r &= r_{Earth} + \rho      
    \end{split}
\end{equation}
where $\rho$ is the altitude above the sea-level of the localization where the sky imager is installed.

\begin{figure}[!htb]
    \begin{subfigure}{0.55\textwidth}
        \centering
        \resizebox{8.125cm}{!}{\begin{tikzpicture}
    \shade[ball color = blue, opacity = 0.25] (0,0) circle (5.cm);
    \shade[ball color = brown, opacity = 0.75] (0,0) circle (3.cm);
    
    \draw (0,0) circle (3 cm);
    \draw (0,0) circle (5 cm);
    \draw [dashed, opacity = 0.4] (-4, 3) -- (4, 3);
    
    \node at (-4, 3) [xshift=-.25cm, yshift=0.cm]{A};
    \node at (4, 3) [xshift=0.25cm, yshift=0.cm]{B};

    \node at (2.25, 4.5) [xshift=.375cm, yshift=0.cm]{$D_i$};
    \node at (-2.25, 4.5) [xshift=-.375cm, yshift=0.cm]{$C_i$};
    \node at (-5, .15) [xshift=-.25cm, yshift=0.cm]{$E_i$};
    
    \draw [dashed, opacity = 0.4] (-2.25, 4.5) -- (2.25, 4.5);
    
    \draw [<->] (0., 4.5) -- (0., 5) node [midway, align = right, xshift=0.25cm, yshift=0.cm]{$\ell_i$};
    
    \draw [<->] (-0.25, 3.05) -- (-0.25, 5) node [midway, align = right, xshift=0.25cm, yshift=0.cm]{$h$};
    
    \draw [<->] (0, 0) -- (3, 0) node [midway, align = right, xshift=0cm, yshift=0.25cm]{$r$};
    
    \draw [<->] (-.15, 3) -- (2.25, 3) node [midway, align = right, xshift=0.35cm, yshift=-0.18cm]{$w_i$};

    \draw (2.25, 4.5) -- (2.25, 3) node [midway, align = right, xshift=0.25cm, yshift=0.cm]{$v_i$};
    
    \draw [<->] (-0.25, 3) -- (2.25, 4.5) node [midway, align = right, xshift=-0.15cm, yshift=0.15cm]{$z_i$};
    
    \draw [<->] (-5, .15) -- (2.25, 4.5) node [midway, align = right, xshift=0cm, yshift=-0.3cm]{$2s_i$};
    
    \draw [<->] (.7, 3) arc (0:30:1) node [midway, align = right, xshift=0.25cm, yshift=0.05cm]{$\varepsilon_i$};
    
    \draw [|<->|] (-2.3, 4.6) arc (117.5:62.5:5)  node [midway, align=right, xshift=0cm, yshift=0.225cm]{$2\hat{y}_{i}$};
        
\end{tikzpicture}}
    \end{subfigure}
    \begin{subfigure}{0.45\textwidth}
        \centering
        \resizebox{6.5625cm}{!}{\begin{tikzpicture}
    \shade[ball color = blue, opacity = 0.25] (0,0) circle (6cm);
    \shade[ball color = brown, opacity = 0.75] (0,0) circle (2.cm);
    
    \draw (0,0) circle (6cm);
    \draw (0,0) circle (2.cm);
    
    \draw [|<->|] (0.25, 5.45) -- (0.25, 6) node [midway, align = right, xshift=0.35cm, yshift=0.cm]{$\lambda_{ij}$};
    
    \node at (0., 6) [xshift=-.225cm, yshift=-.2cm]{$D_i$};
    \node at (0., -6) [xshift=0., yshift=-0.25cm]{$E_i$};
    
    \draw [<->, opacity = 0.8] (0., 6) -- (0., -6) node [midway, align = right, xshift=0.35cm, yshift=0cm]{$2s_i$};
    
    \draw [<->] (-.68, 3.05) arc (110:70:2) node [midway, align = right, xshift=0cm, yshift=-.25cm]{$\alpha_i$};
    
    \draw [|<->|] (-2.675, 5.65) arc (117.5:62.5:5.75)  node [midway, align=right, xshift=0cm, yshift=0.225cm]{$2x^\prime_{ij}$};
       
    \draw [dashed, opacity = 0.4] (-2.5, 5.45) -- (2.5, 5.45);

    \draw [dashed, opacity = 0.4] (0, 1.95) -- (2.5, 5.45);
    
    \draw [dashed, opacity = 0.4] (-2.5, 5.45) -- (0, 1.95);

    \draw [|<->|] (-2.5, 5.25) -- (2.5, 5.25) node [midway, align = right, xshift=0.cm, yshift=-0.3cm]{$2\hat{x}_{ij}$};

\end{tikzpicture}}
    \end{subfigure}
    \caption{Drawing of the \emph{great} circle (surface of a cloud layer) and the \emph{small} circle (Earth's surface). The key in this approach is to find the relation between the chords $C_iD_i$ and $AB$ to calculate $y^\prime_i$ (see right drawing, which is the imager's y-axis view). Similarly, $x^\prime_{i,j}$ is computed for each $y^\prime_i$, using the circle with diameter $2s_i$, formed by chord $D_iE_i$ (see left drawing, which is the imager's x-axis view).}
    \label{fig:great_circle}
\end{figure}

\begin{figure}[!htb]
    \centering
    \resizebox{12.5cm}{!}{\input{cross_section.tikz}}
    \caption{Drawings of the geospatial reprojection in the y-axis or side view (top drawing) and the x-axis or top view (bottom drawing). $Sensor$ and $Lens$ are parts of the sky imager. They are physically separated by a focal length $f$. The distance $z_i$ from the lens to the cloud layer is detailed in the top graph. $x_j^\prime$ and $y_i^\prime$ represent the coordinates of the pixel in the image (see in Eq.~\ref{eq:great_circle_x-axis} and Eq.~\ref{eq:great_circle_y-axis}). When the coordinate system is centered, applying Eq.~\ref{eq:origen}, $\mathbf{x}_0 = \{x_0^\prime, y_0^\prime$\} represents the origin of coordinates.}
    \label{fig:cross_section}
\end{figure}

The imager elevation angle $\varepsilon_i$ defines the triangle formed by the line $z_i$ that intersect the Earth's surface and the cloud layer as:
\begin{equation}
    \tan \varepsilon_i = \frac{v_i}{w_i}.
\end{equation}
By taking this approach, the geospatial reprojection coodiantes are calculated with respect to the imager lens.

\subsection{Reprojection of the y-axis}

The sagitta $\ell_i$ of chord $C_iD_i$ is related to the chord $AB$ (Fig.~\ref{fig:great_circle}). The formula that describes the sagitta $\ell_i$ is a function of the triangle formed by the intersecting line $z_i$ that goes from $AB$ to $C_iD_i$ with elevation angle $\varepsilon_i$,
\begin{equation}
    \begin{split}
        \ell_i &= R - \sqrt{R^2 - w_i^2} \\
        h - v_i &= R - \sqrt{R^2 - w_i^2}\\
        R^2 - w_i^2 &= \left( w_i \tan \varepsilon_i + r \right)^2 \\
        R^2 - w_i^2 &= w_i^2 \tan^2 \varepsilon_i + r^2 + 2r w_i \tan \varepsilon_i  \\
        \left( r + h \right)^2 &= w_i^2 \tan^2 \varepsilon_i + 2 r w_i \tan \varepsilon_i + w_i^2 + r^2 \\
        h^2 + 2 r h &= w_i^2 \left( 1 + \tan^2 \varepsilon_i \right) + 2 r w_i \tan \varepsilon_i \\
        0 &= w_i^2 \left( 1 + \tan^2 \varepsilon_i \right) + w_i \left( 2 r \tan \varepsilon_i \right) - h \left(h + 2 r \right),
    \end{split}
\end{equation}
where $\ell_i = h - v_i$, $v_i = w_i \tan \varepsilon_i$ and  $R = r + h$. The quadratic equation has following coefficients,
\begin{equation}
    \begin{split}
        a_i &= 1 + \tan^2 \varepsilon_i \\
        b_i &= 2 r \tan \varepsilon_i \\
        c_i &= - h \left( h + 2 r \right).
    \end{split}
\end{equation}
The length of triangle side $w_i$ is the result obtained solving the quadratic formula,
\begin{equation}    
    w_i = \frac{- b_i + \sqrt{b_i^2 - 4 a_i c_i}}{2 a_i}, \ w_i \in \mathbb{R}^+.
\end{equation}
When $r \rightarrow \infty$, $w_i \approx d$ and $v_i \approx h$, thus the \emph{flat} approximation is equivalent to the \emph{great} circle approach $w_i \approx h / \tan \varepsilon_i$.

The \emph{great} circle segment $\hat{y}_i$ is the distance from the center of the arc defined by the saggita $\ell_i$ to the point $D_i$ (Fig.~\ref{fig:great_circle}). The chord $2w_i$ is projected to the arc $2\hat{y}_i$ of the \emph{great} circle by applying the arc formula:
\begin{equation}
    \hat{y}_i = R \arcsin \frac{w_i}{R}.
\end{equation}

Each pixel in an image has a different elevation angle $\varepsilon_i$ that corresponds to a point $D_i$ in the \emph{great} circle. Therefore, the coordinates of the pixels relative to the imager lens in the atmosphere cross-section plane are calculated subtracting them the distance $\hat{y}_{sup}$ which has the highest elevation (Fig.~\ref{fig:cross_section}, upper pane),
\begin{equation}\label{eq:great_circle_y-axis}
    y_i^\prime = \hat{y}_i - \hat{y}_{sup}, \quad \forall i = 1, \ \dots, \ N.
\end{equation}

\subsection{Reprojection of the x-axis}

The reprojection of the sensor plane x-axis to the atmosphere cross-section is a function of the distance $z^2_i = w_i^2 + v_i^2$ from the sensor plane to the cloud layer, and the chord $2\hat{x}_{i,j}$ of segment the $2x^\prime_{i,j}$ formed by the angle $\alpha_j$ (Fig.~\ref{fig:cross_section}, lower pane),
\begin{equation}    
    \hat{x}_{i,j} = \left( z_i - \lambda_{i,j} \right) \tan \alpha_j, \quad \forall i = 1, \dots, M, \ j = 1, \dots, M.
\end{equation}
The diameter of the small circle $2s_i$, which is the chord $D_i E_i$ in Fig.~\ref{fig:great_circle}, is obtained by applying the intersecting chord theorem. In Euclid's Elements Book III, Proposition 35, (see \cite{heath1956}), the intersecting chords theorem is defined as $|AS| \cdot |SC| = |BS| \cdot |SD| = r^2 - d^2$. When this theorem is applied to our problem the corresponding variables are $d = (R - h)$, $r = R$, $|AS| = s_i - z_i$ and $|SC| = z_i$ (see Fig.~\ref{fig:great_circle} y-axis graph), so
\begin{equation}
    \begin{split}
        \left(2s_i - z_i \right) z_i &= R^2 - (R - h)^2 \\
        s_i &= \frac{2Rh - h^2}{2z_i} + \frac{z_i}{2}.  
    \end{split}
\end{equation}
The relation between the arc length $2x^\prime_{i,j}$ and the chord $2\hat{x}_{i,j}$ is found through the sagitta $\lambda_{i,j}$. The formula which describes \begin{equation}\label{eq:quadratic_equation_x}
    \begin{split}
        \left( 2 s_{i} - \lambda_{i,j} \right) \lambda_{i,j} &= \hat{x}_{i,j}^2 \\
        2 s_{i} &= \lambda_{i,j} + \frac{\hat{x}_{i,j}^2}{\lambda_{i,j}} \\
        2 s_i \lambda_{i,j} - \lambda^2_{i,j} &= \left( z_i - \lambda_{i,j} \right)^2 \tan^2 \alpha_j \\
        0 &= \lambda^2_{i,j} \left( 1 + \tan^2 \alpha_j \right) - 2 \lambda_{i,j} \left( z_i \tan^2 \alpha_j - s_i \right) + z_i^2 \tan^2 \alpha_j,
    \end{split}
\end{equation}
where coefficients for solving the quadratic formula are,
\begin{equation}
    \begin{split}
        a_{i,j} &= 1 + \tan^2\alpha_j \\
        b_{i,j} &= - 2 s_i - 2 z_i\tan^2\alpha_j  \\
        c_{i,j} &= z_i^2 \tan^2\alpha_j.
    \end{split}
\end{equation}
The sagitta $\lambda_{i,j}$ is the result obtained solving Eq.~\eqref{eq:quadratic_equation_x},
\begin{equation}    
    \lambda_{i,j} = \frac{- b_{i,j} - \sqrt{b_{i,j}^2 - 4  a_{i,j}  c_{i,j}}}{2  a_{i,j}}, \quad \lambda_{i,j} \in \mathbb{R}^+.
\end{equation}
When $r \rightarrow \infty$, $s_i \rightarrow \infty$, in consequence $\lambda_{i,j} \approx 0$ and $x^\prime_{i,j} \approx \hat{x}_{i,j}$. The \emph{flat} Earth approximation is equivalent to the \emph{great} circle approach.

The arc length $2 x_{i,j}^\prime$ is calculated knowing the sagitta $\lambda_{i,j}$ and the small radius $s_i$,
\begin{equation}\label{eq:great_circle_x-axis} 
    x_{i, j}^\prime = r^\prime_{i} \arcsin \left[ \frac{\left(z_i - \lambda_{i,j}\right) \tan \alpha_j}{r^\prime_{i}} \right]
\end{equation}
where segment $x_{i, j}^\prime$ is the projection of x-axis in the atmosphere cross-selection plane.

The origin of the coordinate system can be defined at the position of the Sun,
\begin{equation}\label{eq:origen}
    \begin{split}
         x_{i,j}^{\prime\prime} &= x_{i,j}^{\prime} - x^{\prime}_0 \\
         y_{i,j}^{\prime\prime} &= y_{i,j}^{\prime} - y^{\prime}_0,
    \end{split}  
\end{equation}
where $\mathbf{x}^\prime_0 = \{y^\prime_0, x^\prime_0\}$ are the pixel index of the Sun position in the image. These equation is applicable to both proposed perspective reprojections.

\section{Results and Discussion}

The geospatial perspective reprojections are applied to a sky imager mounted on a solar tracker that updates its pan and tilt every second, maintaining the Sun in a central position in the images throughout the day. The sky imager is located in the Electrical and Computer Engineering (ECE) department at the University of New Mexico (UNM) central campus in Albuquerque. The climate of Albuquerque is arid semi-continental, with minimal rain, which is more likely in the summer months. The ECE department is approximately located $1,250$m away (i.e., linear distance) from the city center whose elevation is $1,620$m with respect to sea level.

The IR sensor is a Lepton\footnote{https://www.flir.com/} 2.5 radiometric camera with wavelength from 8 to 14$\mu$m. Pixel intensity within the frame is measured in centikelvin units. The resolution of an IR image is $80 \times 60$ pixels. To implement the reprojection, the manufacturing specifications of the camera used are: $ 63.75^\circ$ diagonal $\mathrm{FOV}$,  $51^\circ$ horizontal $\mathrm{FOV}_x$, $38.25^\circ$ vertical $\mathrm{FOV}_y$, and the size of a pixel is $ \delta = 17\mu$m. When other lenses (e.g. fisheye) are used, the camera lens affine reprojection must first be computed to know the FOV of each pixel.

\begin{figure}[!htb]
    \begin{subfigure}{\linewidth}
        \centering
        \includegraphics[scale = 0.2, trim = {4.5cm, 0cm, 4.5cm, 0cm}, clip]{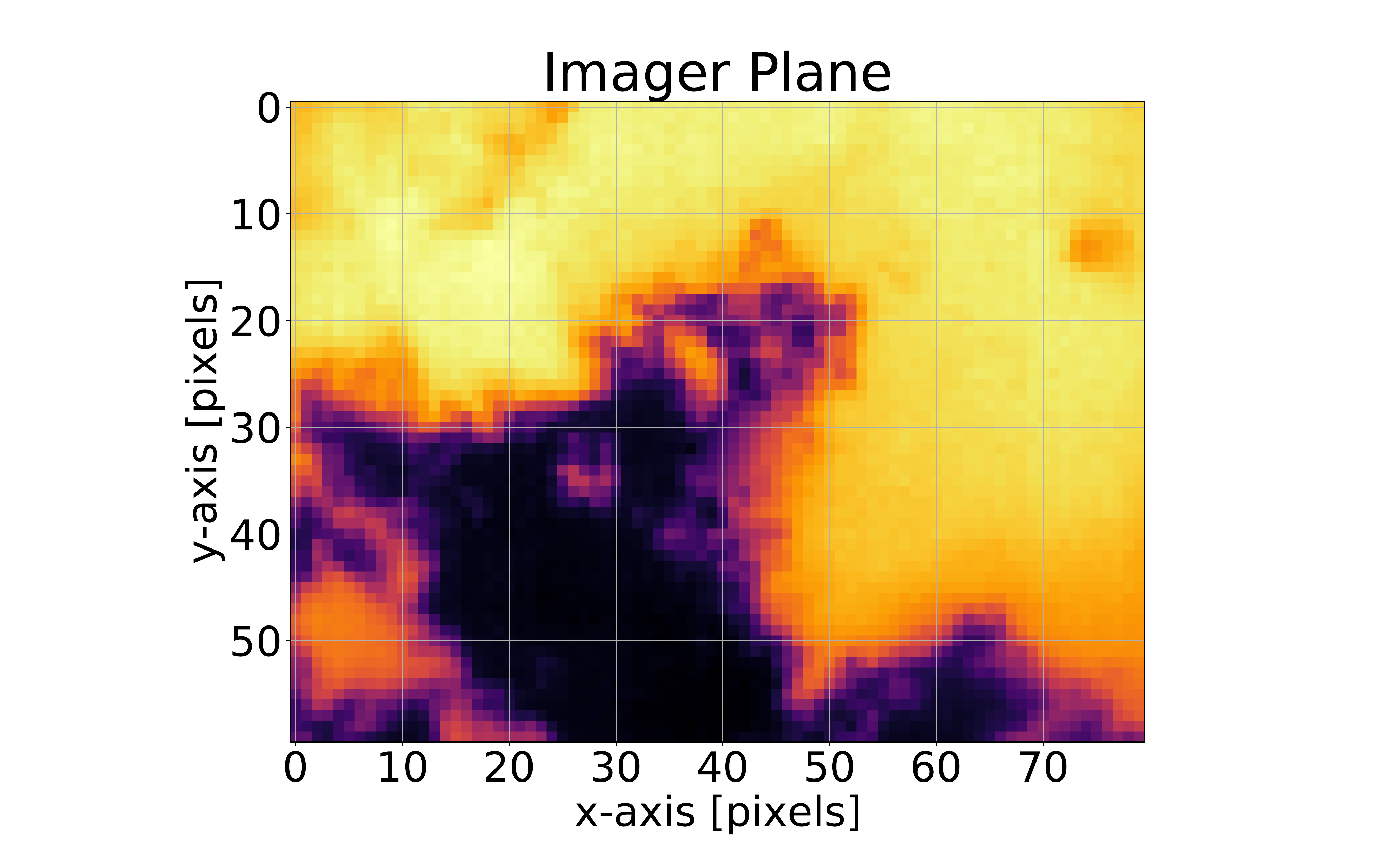}
        \includegraphics[scale = 0.2]{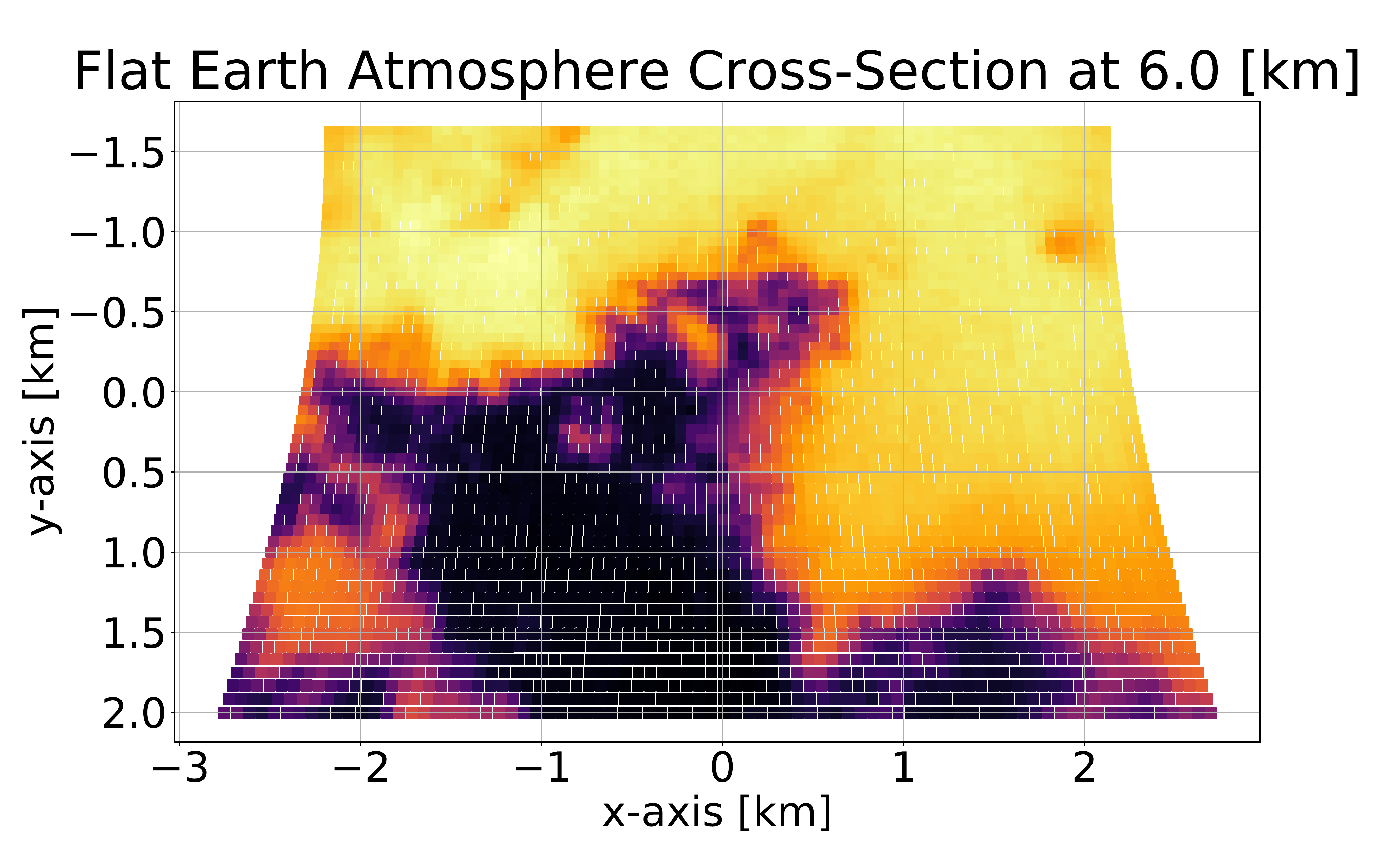}
    \end{subfigure}
    \begin{subfigure}{\linewidth}
        \centering
        \includegraphics[scale = 0.2, trim = {4.5cm, 0cm, 4.5cm, 0cm}, clip]{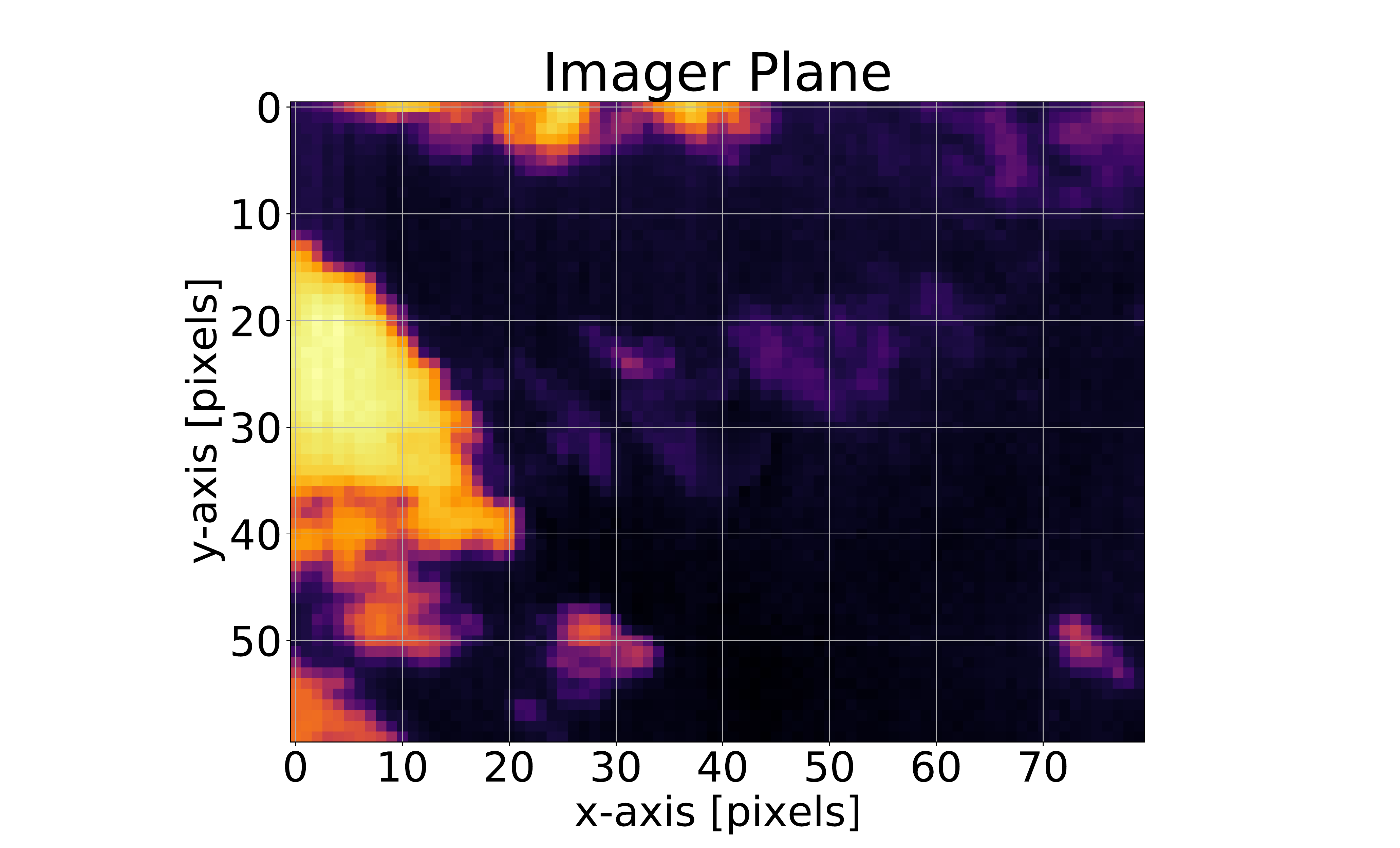}
        \includegraphics[scale = 0.2]{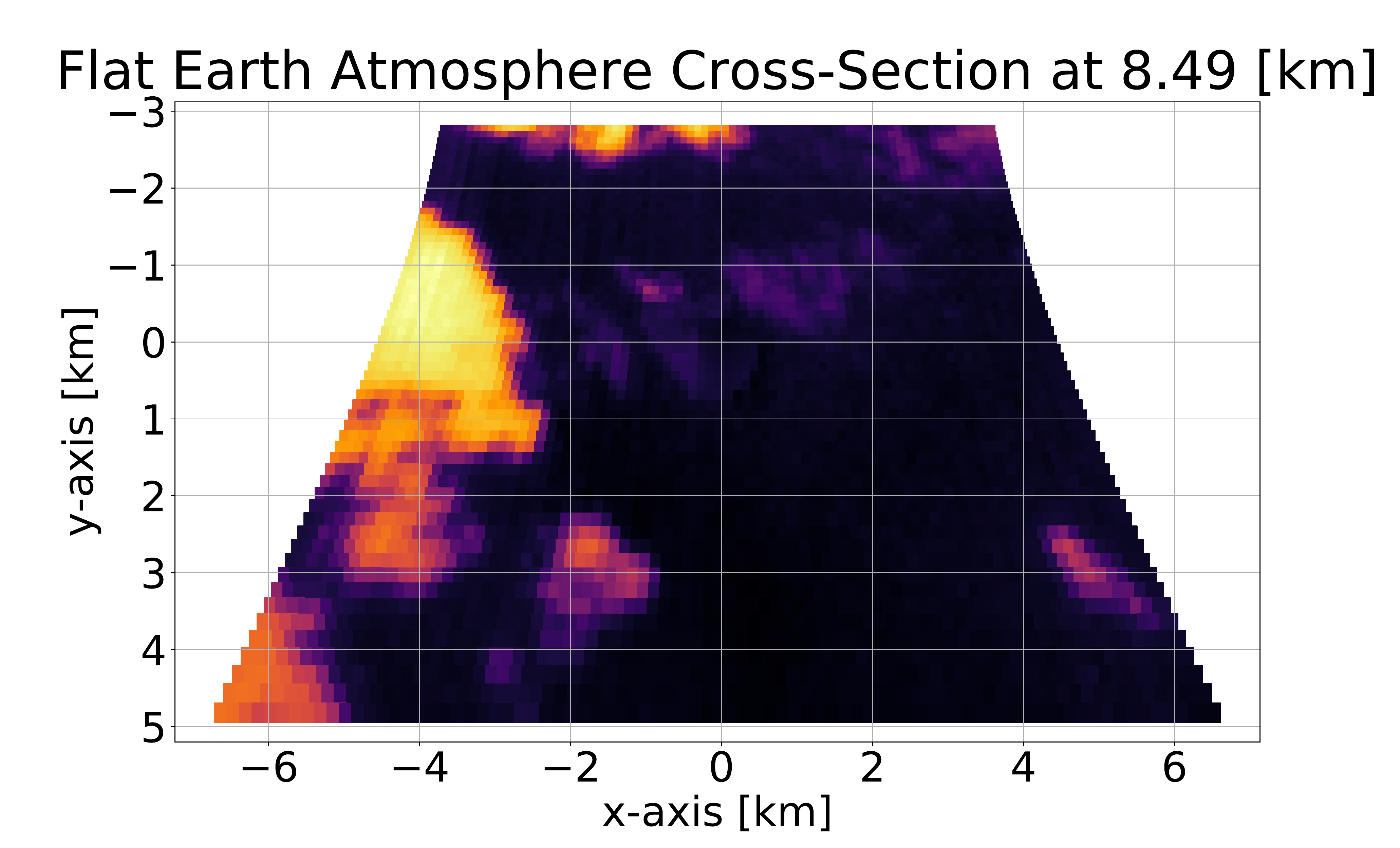}
    \end{subfigure}
    \begin{subfigure}{\linewidth}
        \centering
        \includegraphics[scale = 0.2, trim = {4.5cm, 0cm, 4.5cm, 0cm}, clip]{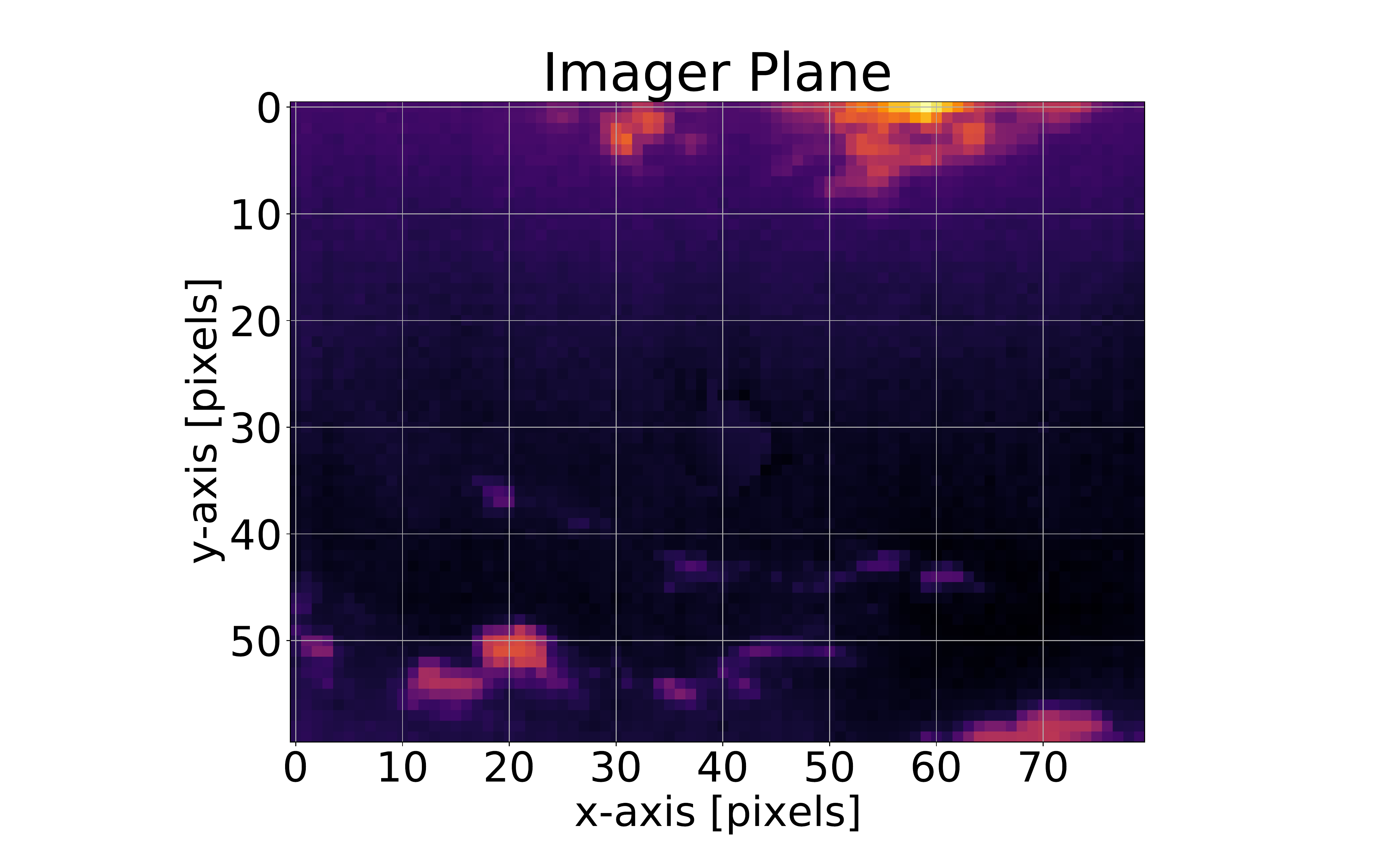}
        \includegraphics[scale = 0.2]{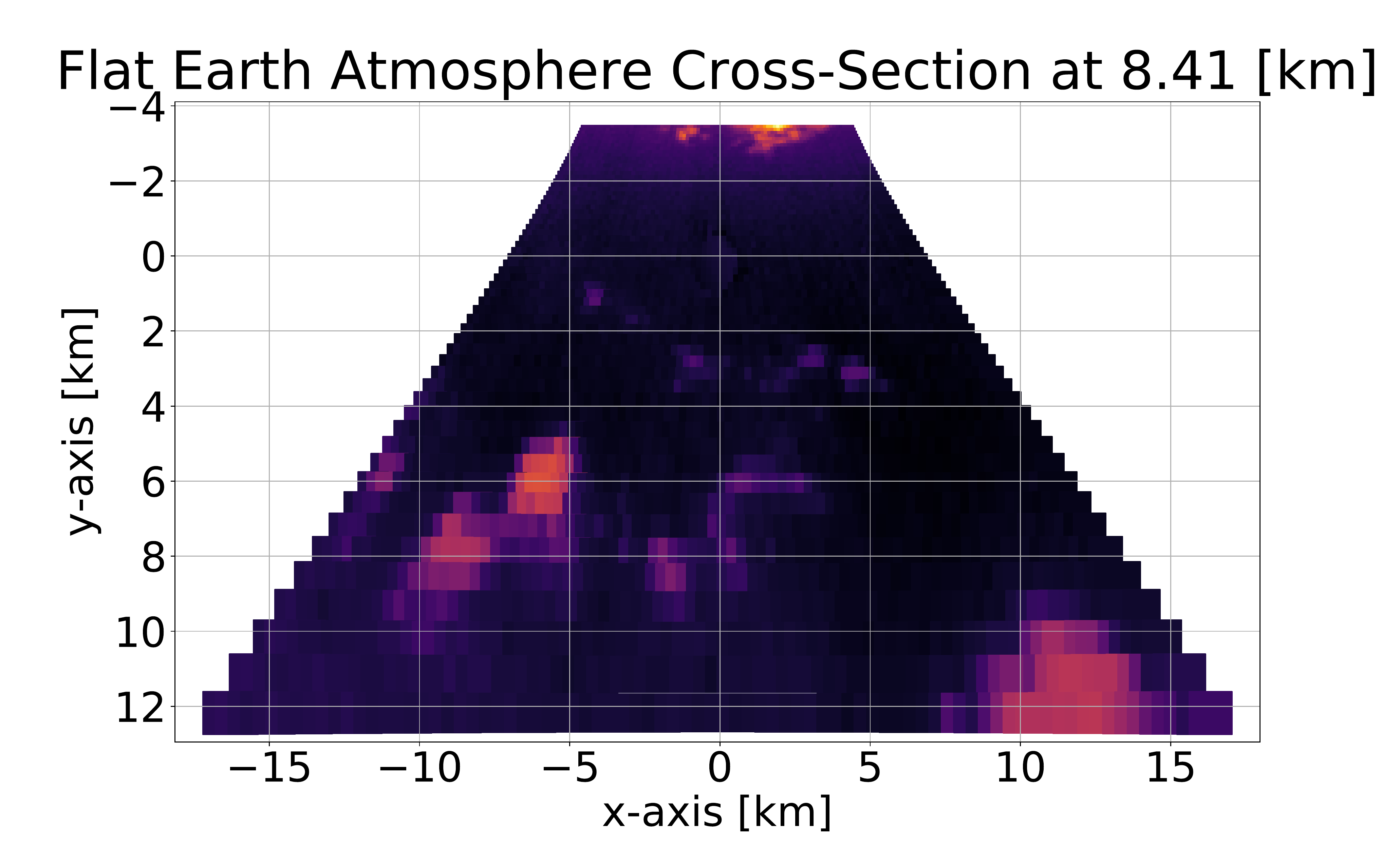}
    \end{subfigure}
    \caption{The left column shows three sky images taken at different elevation angles: $71.06^\circ$, $50.17^\circ$ and $30.83^\circ$ (from top to bottom). The right column shows the same three images after applying the geospatial perspective reprojection using the \emph{flat} Earth approximation.}
    \label{fig:flat-earth_cross-section}
\end{figure}

\begin{figure}[!htb]
   \begin{subfigure}{\linewidth}
        \centering
        \includegraphics[scale = 0.2, trim = {4.5cm, 0cm, 4.5cm, 0cm}, clip]{images/imager_71.0.pdf}
        \includegraphics[scale = 0.2, trim = {0cm, 0cm, 0cm, 0cm}, clip]{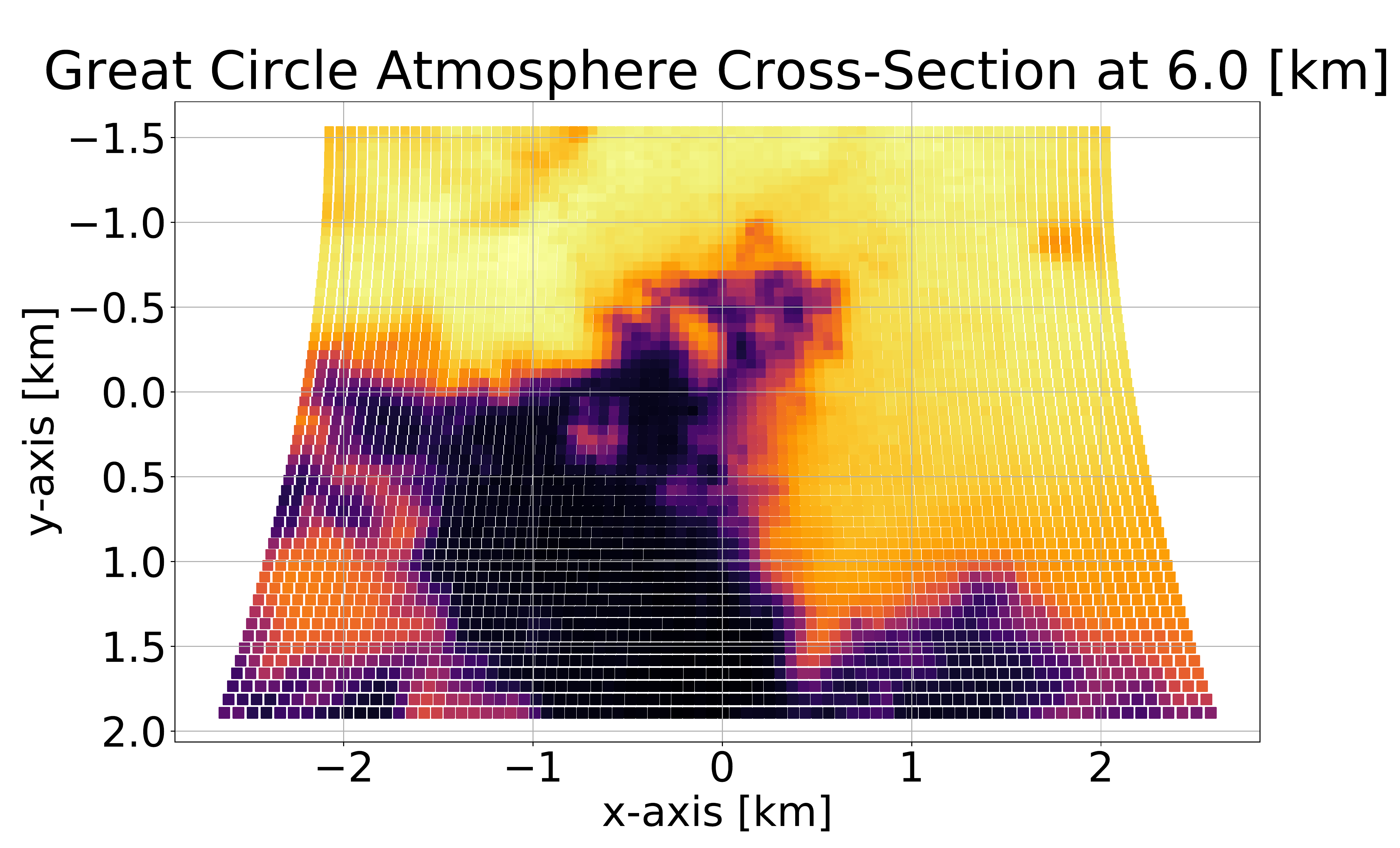}
    \end{subfigure}
    \begin{subfigure}{\linewidth}
        \centering
        \includegraphics[scale = 0.2, trim = {4.5cm, 0cm, 4.5cm, 0cm}, clip]{images/imager_50.0.pdf}
        \includegraphics[scale = 0.2, trim = {0cm, 0cm, 0cm, 0cm}, clip]{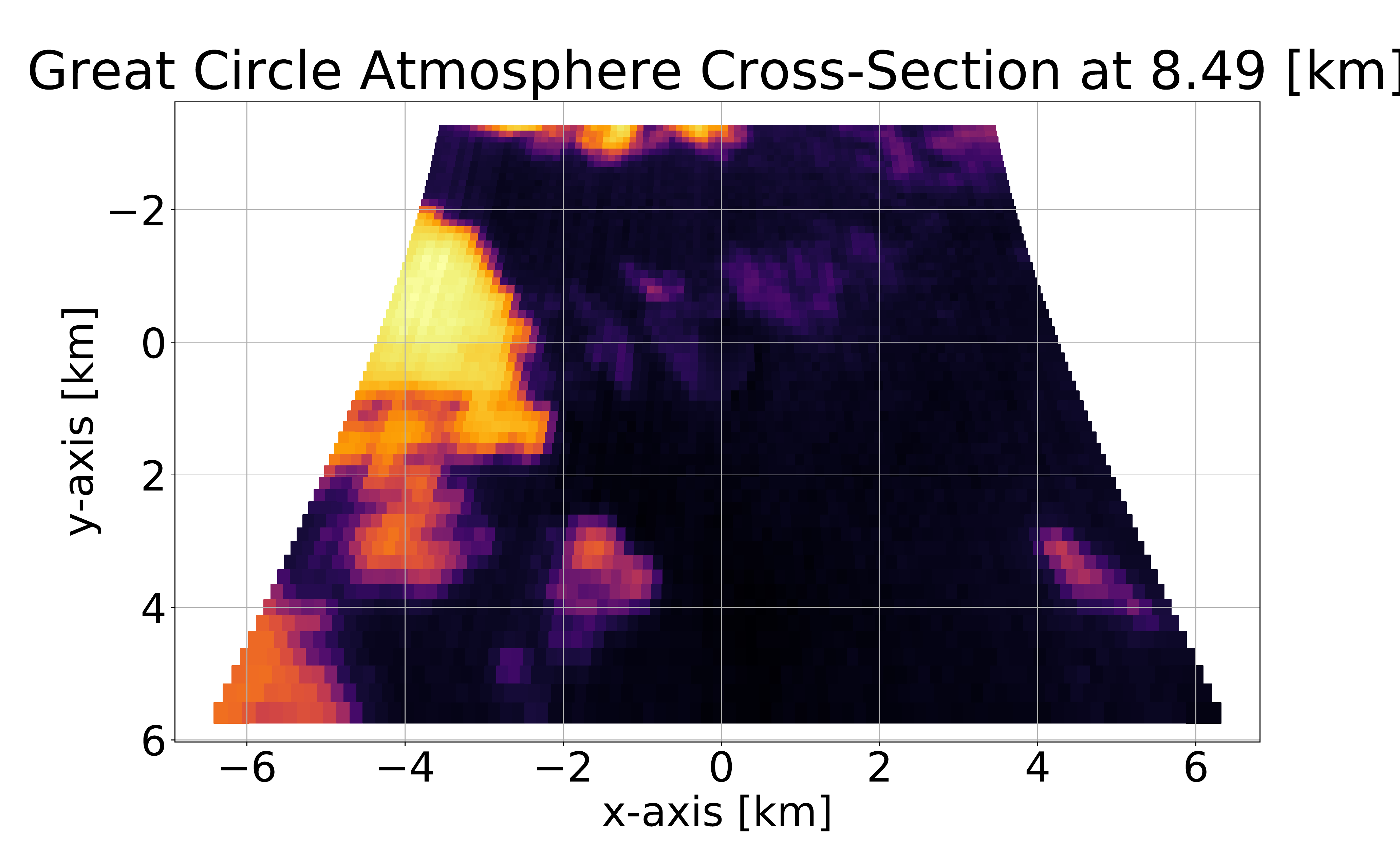}
    \end{subfigure}
    \begin{subfigure}{\linewidth}
        \centering
        \includegraphics[scale = 0.2, trim = {4.5cm, 0cm, 4.5cm, 0cm}, clip]{images/imager_31.0.pdf}
        \includegraphics[scale = 0.2, trim = {0cm, 0cm, 0cm, 0cm}, clip]{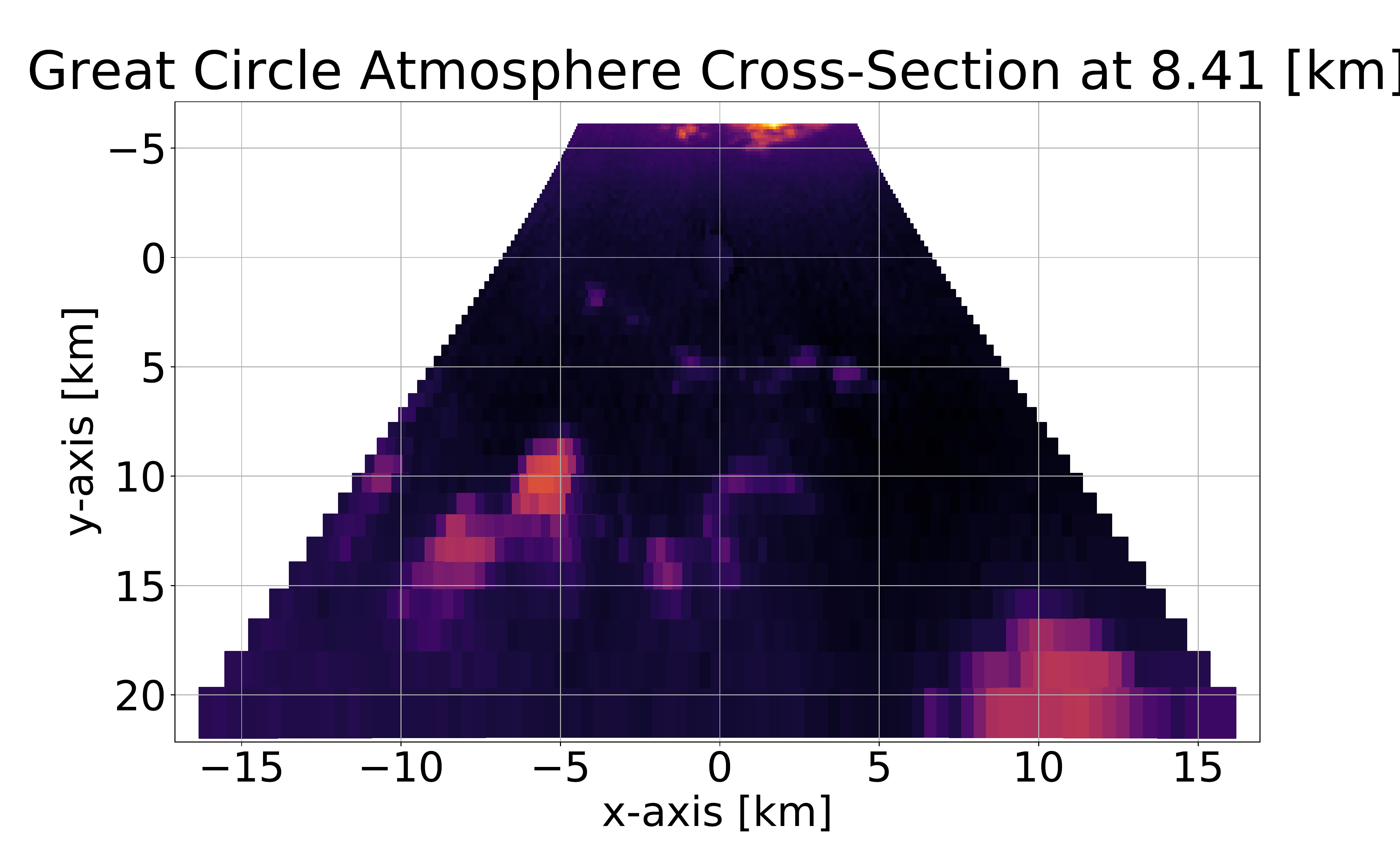}
    \end{subfigure}
    \caption{The left column shows three sky images taken at different elevation angles: $71.06^\circ$, $50.17^\circ$ and $30.83^\circ$ (from top to bottom). The right column shows the resulting sky images after applying the geospatial perspective reprojection using the \emph{great} circle approach.}
    \label{fig:great-circle_cross-section}
\end{figure}

The pixels in Fig.~\ref{fig:flat-earth_cross-section} and \ref{fig:great-circle_cross-section} are displayed in the camera pixel coordinates (left) and in the atmosphere cross-section plane (right). The pixels are scaled to their actual size in the atmosphere cross-section plane. The distortion produced by the sky imager perspective causes the atmosphere cross-section plane dimensions to increase when the elevation angle decreases.

The difference between both geospatial reprojections is measured using Root Mean Square Error (RMSE). The coordinates computed using the \emph{flat} Earth assumption are $\mathbf{X}_1, \mathbf{Y}_1$, and the coordinates computed using \emph{great} circle approach are $\mathbf{X}_2,\mathbf{Y}_2$. The RMSE, defined as $\mathbf{E}$, is calculated for each pixel averaging together the difference residuals computed independently in coordinates x and y, 
\begin{equation}
    \mathbf{E} = \sqrt{ \frac{1}{2} \left[ \mathcal{R} \left(\mathbf{X}_1, \mathbf{X}_2 \right) + \mathcal{R} \left(\mathbf{Y}_1, \mathbf{Y}_2 \right) \right] }.
    \label{eq:RMSE}
\end{equation}
The residuals are $\mathcal{R} \left(\mathbf{X}_1, \mathbf{X}_2 \right) = ( \mathbf{X}_1^\prime - \mathbf{X}_2^\prime )^2$ for each coordinate.

\begin{figure}[!htb]
    \begin{subfigure}{\linewidth}
        \centering
        \includegraphics[scale = 0.3625, trim = {.75cm, 0cm, 1.75cm, 0cm}, clip]{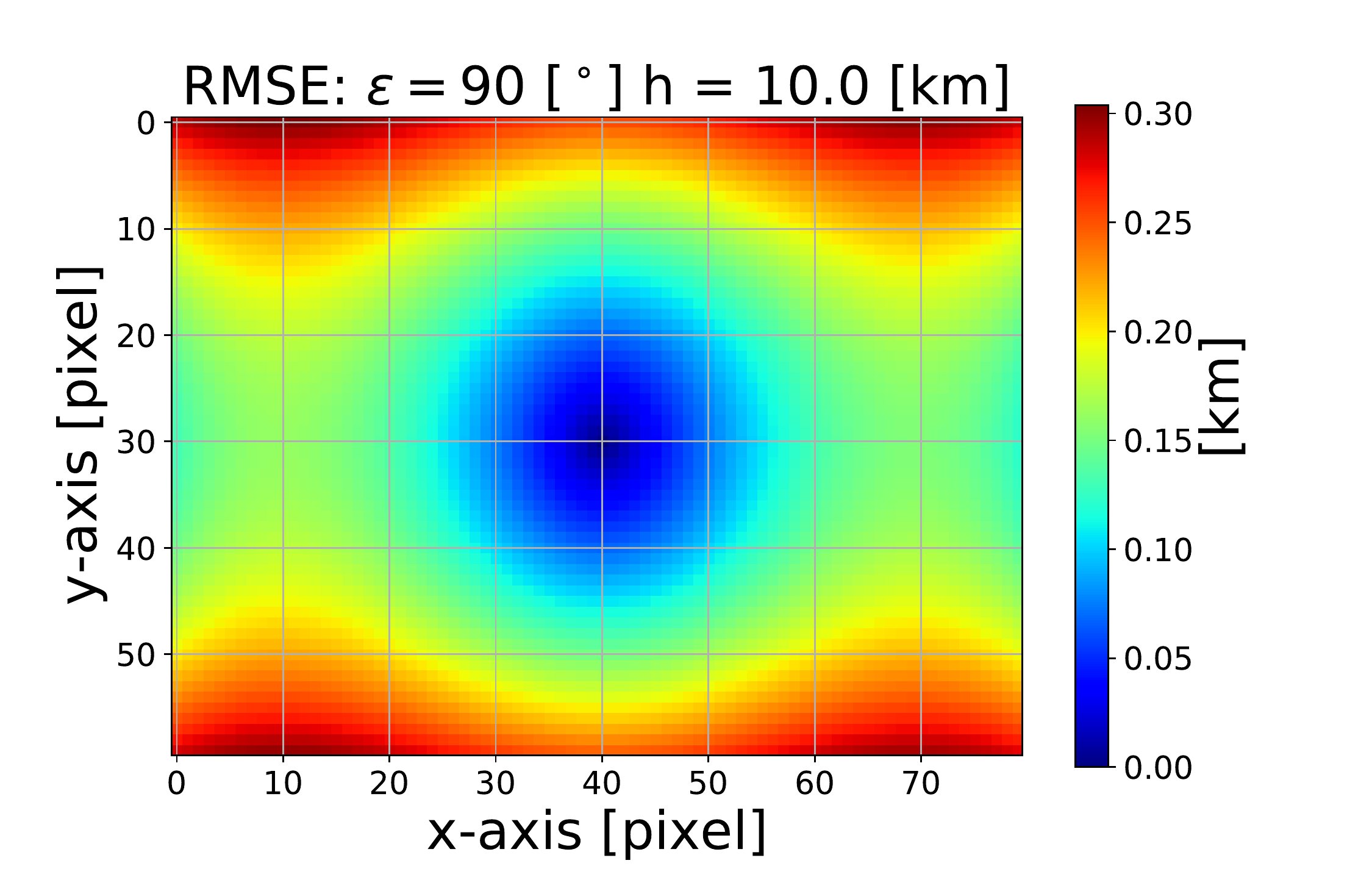}
        \includegraphics[scale = 0.3625, trim = {.75cm, 0cm, 1.75cm, 0cm}, clip]{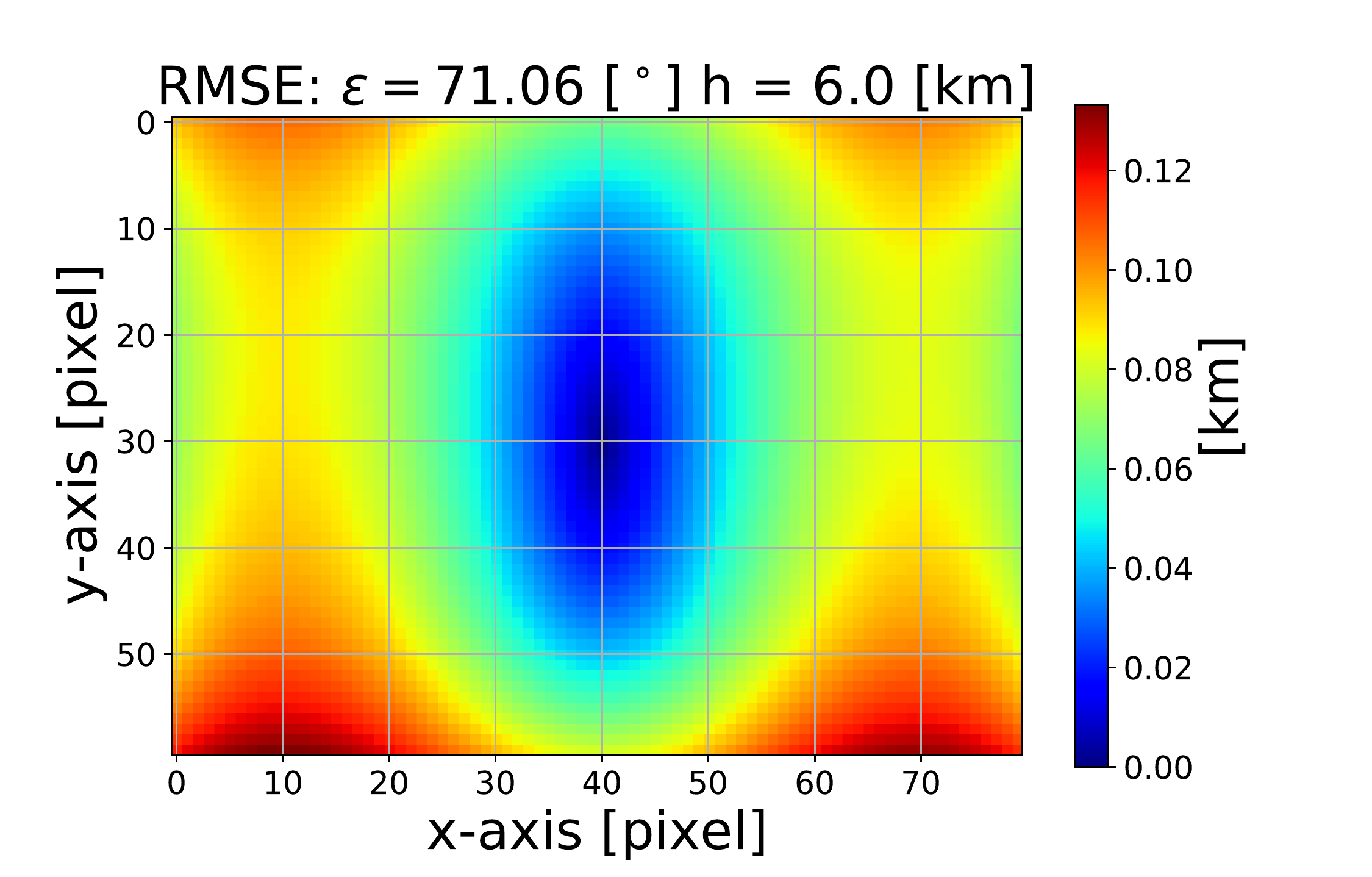}
    \end{subfigure}
    \begin{subfigure}{\linewidth}
        \centering
        \includegraphics[scale = 0.3625, trim = {.75cm, 0cm, 1.75cm, 0cm}, clip]{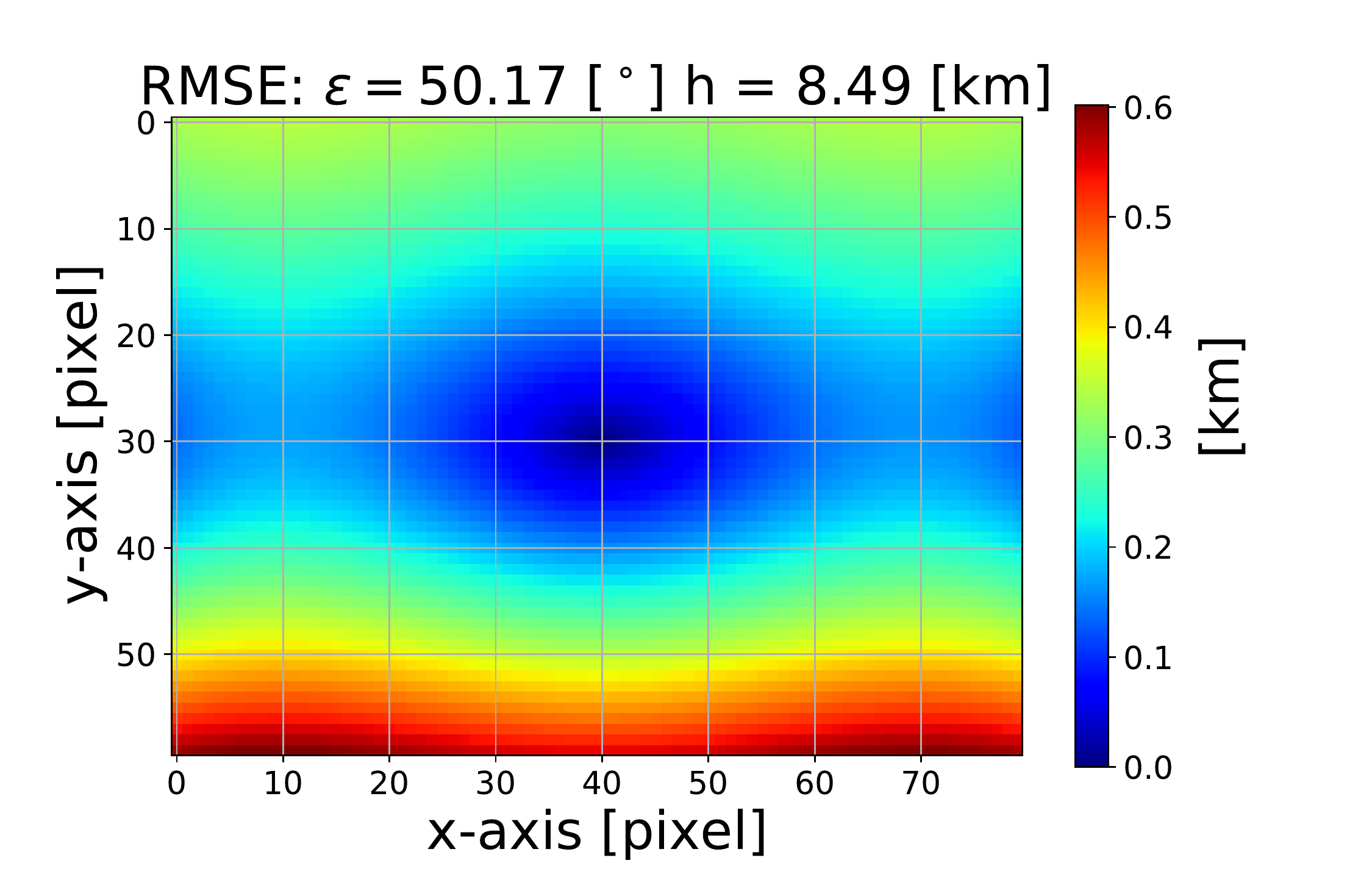}
        \includegraphics[scale = 0.3625, trim = {.75cm, 0cm, 1.75cm, 0cm}, clip]{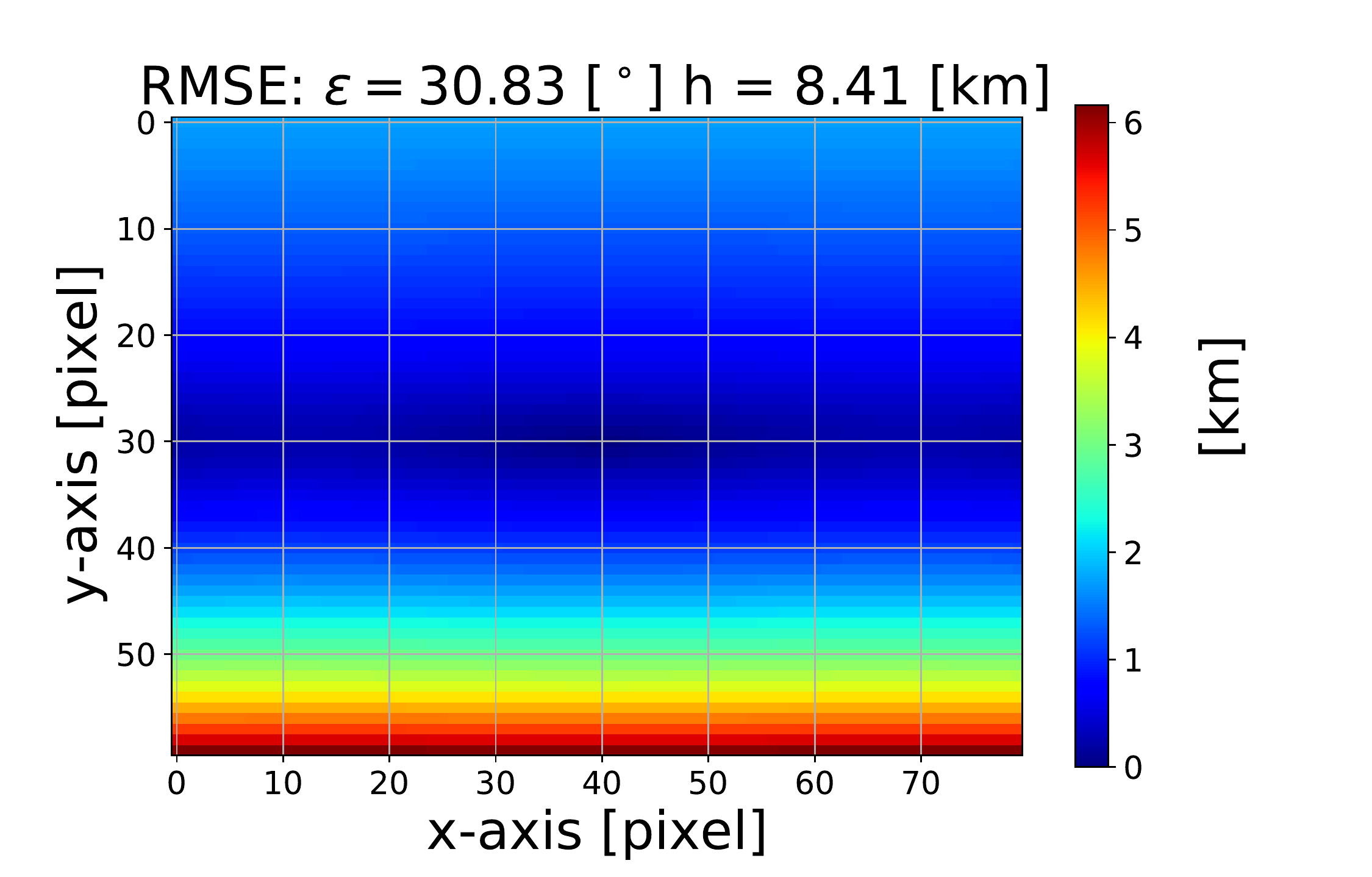}
    \end{subfigure}
    \caption{RMSE between the atmosphere cross-section coordinates approximated using the \emph{flat} Earth assumption reprojection and the \emph{great} circle approach reprojection. The coordinates of each reprojection are displayed in Fig.~\ref{fig:flat-earth_cross-section} and \ref{fig:great-circle_cross-section} respectively.}
    \label{fig:error_maps}
\end{figure}

The error maps (see Fig.~\ref{fig:error_maps}) show the differences between the coordinate systems approximated by both reprojections. The symmetry between both reprojections is not perfectly circular. This is because the elevation angle in \emph{flat} Earth reprojection, was approximated as constant across the pixels in the same row. 

The tropopause average height is approximately $10$km in the latitude where the sky imager is located depending on the season. The first image in Fig.~\ref{fig:error_maps} shows the error map when the camera is at the zenith. The magnitude order of the error is in meters when $\varepsilon \geq 30^\circ$. However, when the Sun's elevation angle is below $\varepsilon < 30^\circ$ the magnitude order of the error is in kilometers. Taking this into account, the geospatial reprojection that assumes that the Earth’s surface is \emph{flat}, is only adequate when the elevation angle of a sky imager pixel is above $\varepsilon \geq 30^\circ$. If the sky imager is designed to operate below $\varepsilon < 30^\circ$, the most suitable reprojection is computed using the great circle approach.

\begin{figure}[!htb]
    \centering
    \includegraphics[scale = 0.35, trim = {0cm, 0cm, 2.5cm, 0cm}, clip]{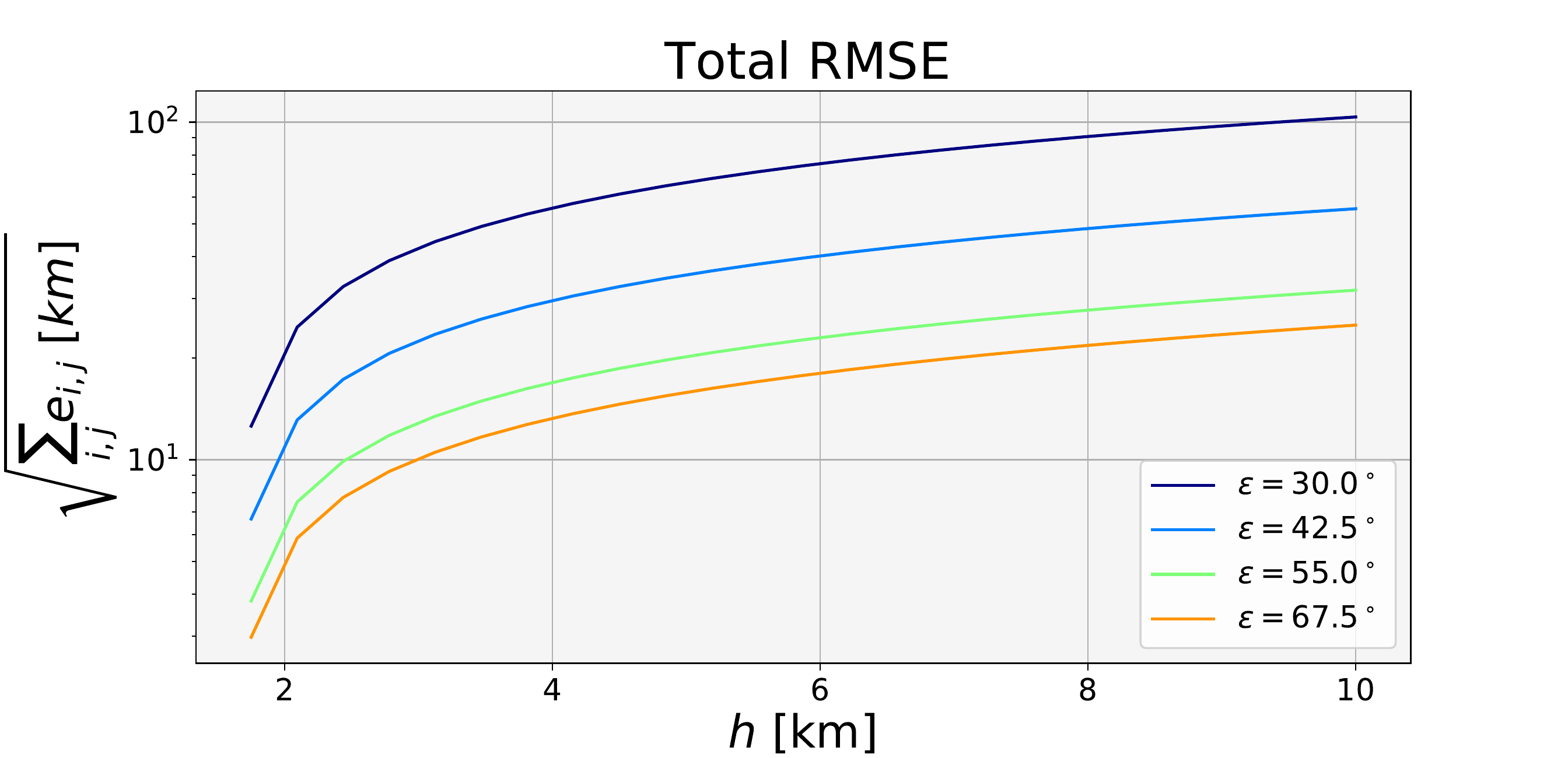}
    \includegraphics[scale = 0.35, trim = {1.5cm, 0cm, 0cm, 0cm}, clip]{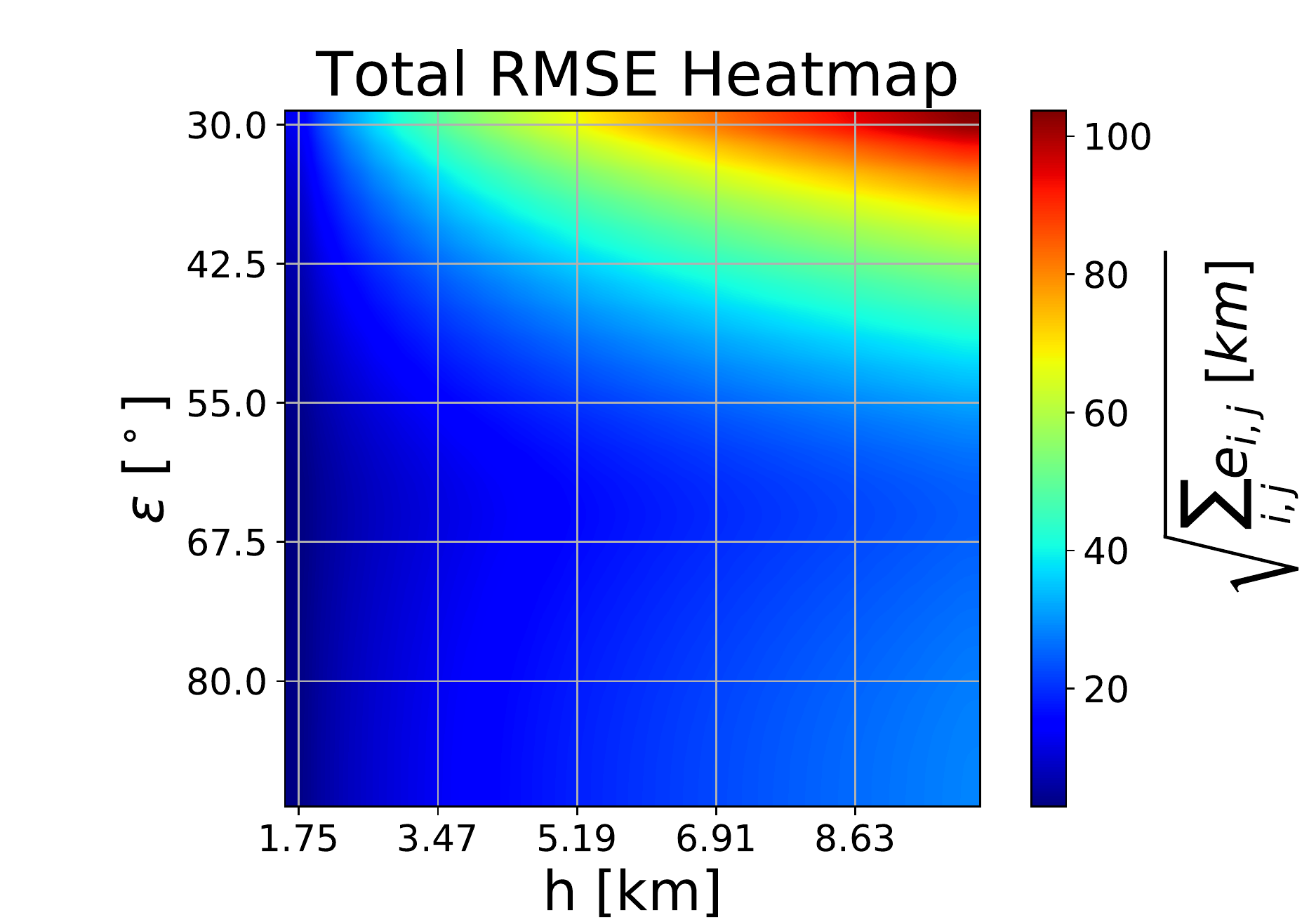}
    \caption{The left graph shows the increase in the quadratic total sum of error as a function of the height for $5$ different elevation angles: $30^\circ$, $42.5^\circ$, $55^\circ$ and $67.5^\circ$. The error function is in Eq.~\eqref{eq:RMSE}. The right graph shows the quadratic total sum of errors as a function of the elevation angle and the height.}
    \label{fig:total_error_maps}
\end{figure}

The difference between both transformations in the magnitude of the error is due to the dimensions of the region of the atmosphere that is being measured with the sky imager (see Fig.~\ref{fig:total_error_maps}). As the elevation angle decreases, the region of the atmosphere that is measured in each pixel increases (i.e., perspective). Consequently, the great circle approach performs a more accurate approximation of the cross-section plane of the atmosphere in which the clouds are moving.      

\begin{figure}[!htb]
    \centering
    \includegraphics[scale = 0.1825, trim = {0.1cm, 0cm, 2.6cm, 0cm}, clip]{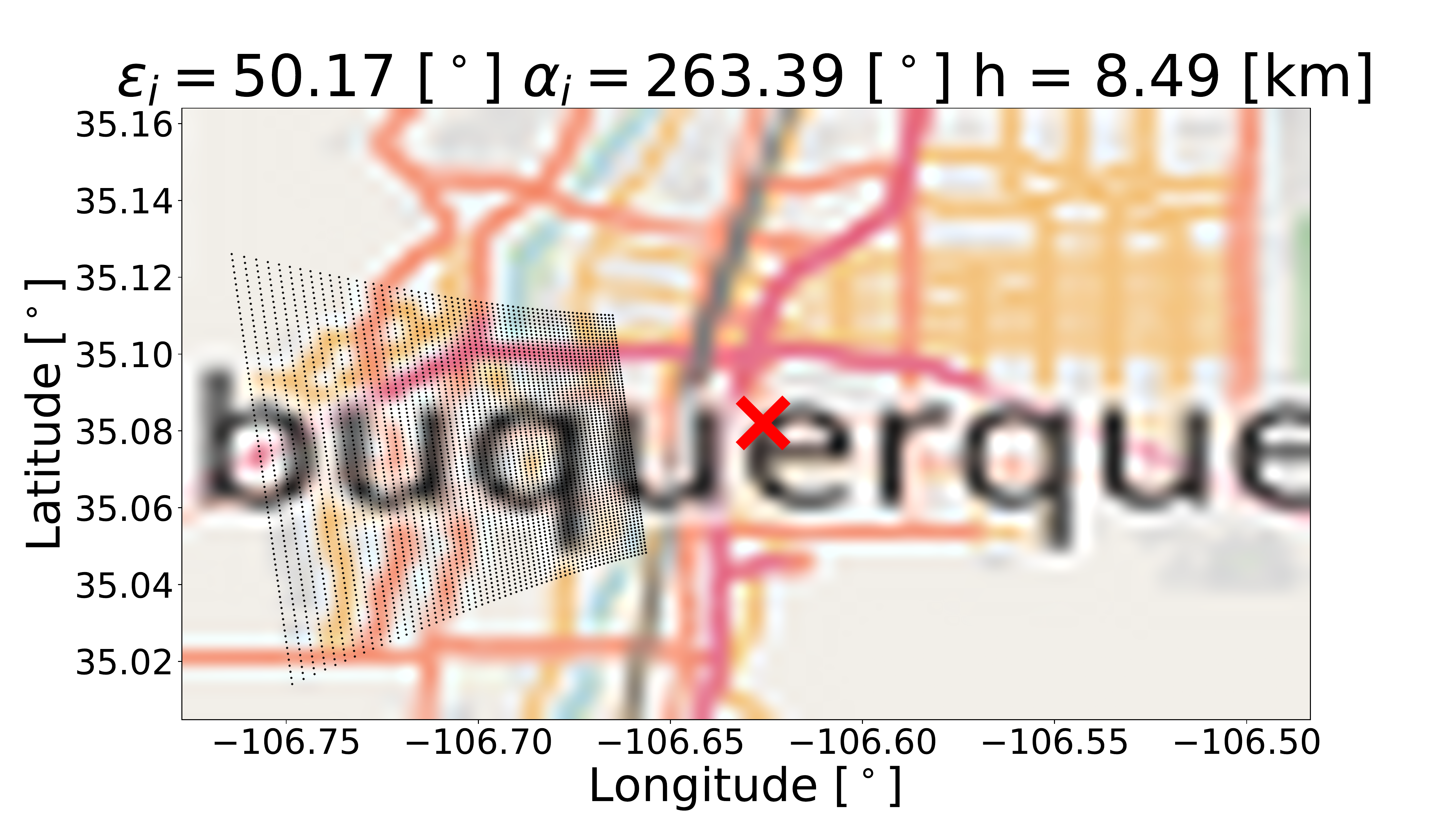}
    \includegraphics[scale = 0.1825, trim = {1cm, 0cm, 2.6cm, 0cm}, clip]{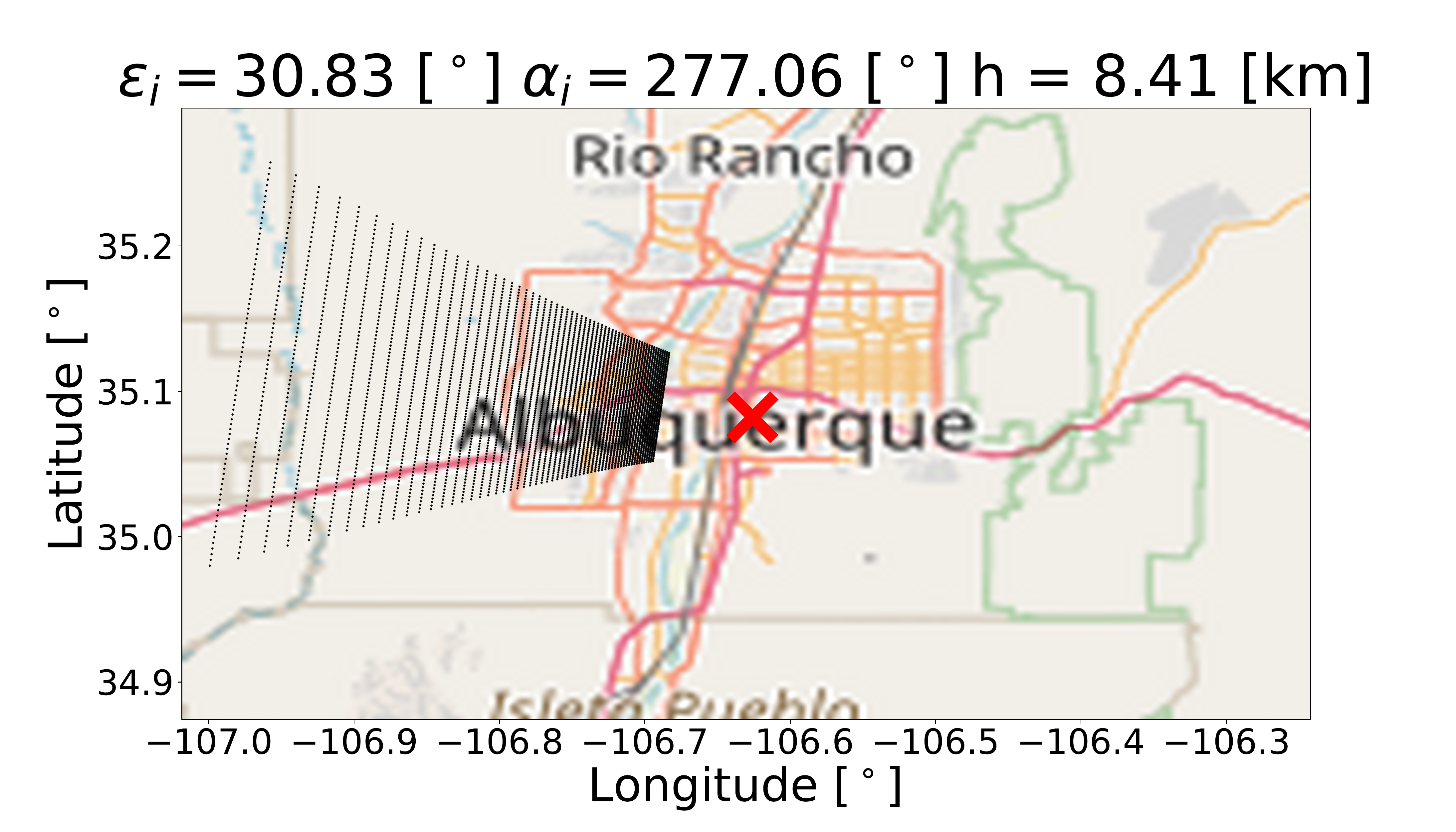}
    \caption{Atmosphere cross-section plane projected on the Earth’s surface for elevations of $50^\circ$ and $31^\circ$. The sky imager localization is the red dot. The sky imager pixels are in black. The coordinates of a pixel are defined by a longitude and latitude angle.}
    \label{fig:ground_projection}
\end{figure}

The atmosphere cross-section projected on the Earth’s surface using the \emph{great} circle approach reprojection is shown in Fig.~\ref{fig:ground_projection} in Geographic Coordinates System (GCS). The GCS components are longitude and latitude and they are defined in degrees. The atmosphere cross-section plane is considerably larger when the Sun's elevation angle is low. The distance between pixels in an image increases exponentially from top to bottom. 

The results presented in this investigation show that the proposed methodology is advantageous with respect to other methods available in the literature from a theoretical and technological point of view. The geometric reprojection proposed by \cite{nouri2019} (i.e., voxel carving) is equivalent to the \emph{flat} Earth approximation investigated in this research, and thus it does not consider the curvature of the Earth (i.e., \emph{great} circle approach). As it is demonstrated in this research (see Fig.~\ref{fig:error_maps}), the order of magnitude of the error produced by this approximation is in the range of kilometers for high clouds (e.g., stratus) measured by pixels with elevation angles $< 30^\circ$. In addition, low-cost radiometric far infrared cameras provide temperature measurements (e.g., \cite{lewis2010}) which can be transformed to height measurements \cite{stone1979} when combined with weather features measured by a simple weather station in the ground \cite{terren2021e}. Radiometric infrared cameras have low resolution \cite{TERREN2020d}, but their resolution is sufficient to perform accurate intra-hour solar forecasting \cite{terren2021f}.

\section{Conclusion}

Intra-hour solar forecasting algorithms utilize consecutive sky images to compute cloud velocity vectors, anticipating when a cloud will occlude the Sun and produce a decrease in the global solar irradiance that reaches the Earth's surface. Velocity vectors are calculated in units of pixels per frame, but the dimensions of the pixels in sky images vary with the elevation angle. Therefore, the velocity vector accuracy used to forecast cloud occlusions of the Sun can be improved. The proposed perspective reprojection of the sensor plane to the geospatial atmosphere cross-section plane can be used to transform the pixels in sky images to the cross section coordinate system of the clouds.

When used in sky imagers, thermal images are advantageous in that cloud height can be approximated when cloud temperature is known. Radiometric infrared cameras composed of microbolometers are an inexpensive technology capable of acquiring thermal sky images. When intersected by the sky-imaging system field of view, the dimensions of the atmosphere cross-section plane can be determined using the proposed reprojections and temperatures of the objects in the images.

\section{Data Availability}

The procedure to acquire and preprocessing the radiometric far infrared sky images, plus the hardware was described in \cite{TERREN2020d}. The data used in this work is publicly available in a DRYAD repository (\url{https://doi.org/10.5061/dryad.zcrjdfn9m}. The software for both geospatial perspective reprojections is available in a GitHub repository (\url{https://github.com/gterren/geospatial_perspective_reprojection}).

\section*{Acknowledgments}

This work has been supported by National Science Foundation (NSF) EPSCoR grant number OIA-1757207 and the King Felipe VI endowed Chair of the UNM. Authors would like to thank the UNM Center for Advanced Research Computing (CARC), supported in part by NSF, for providing the high performance computing and large-scale storage resources used in this work. We would also like to thank Marie R. Fernandez for proof reading the manuscript.

\bibliographystyle{unsrt}  
\bibliography{mybibfile}

\begin{thebibliography}{10}

\bibitem{TZOUMANIKAS2016}
P.~Tzoumanikas, E.~Nikitidou, A.F. Bais, and A.~Kazantzidis.
\newblock The effect of clouds on surface solar irradiance, based on data from
  an all-sky imaging system.
\newblock {\em Renewable Energy}, 95:314 -- 322, 2016.

\bibitem{KONG2020}
Weicong Kong, Youwei Jia, Zhao~Yang Dong, Ke~Meng, and Songjian Chai.
\newblock Hybrid approaches based on deep whole-sky-image learning to
  photovoltaic generation forecasting.
\newblock {\em Applied Energy}, 280:115875, 2020.

\bibitem{MAMMOLI2013}
A.~{Mammoli}, A.~{Ellis}, A.~{Menicucci}, S.~{Willard}, T.~{Caudell}, and
  J.~{Simmins}.
\newblock Low-cost solar micro-forecasts for pv smoothing.
\newblock In {\em 2013 1st IEEE Conference on Technologies for Sustainability
  (SusTech)}, pages 238--243, 2013.

\bibitem{CHU2016}
Yinghao Chu, Mengying Li, and Carlos~F.M. Coimbra.
\newblock Sun-tracking imaging system for intra-hour dni forecasts.
\newblock {\em Renewable Energy}, 96:792 -- 799, 2016.

\bibitem{TERREN2020c}
Guillermo Terrén-Serrano and Manel Martínez-Ramón.
\newblock Comparative analysis of methods for cloud segmentation in
  ground-based infrared images.
\newblock {\em Renewable Energy}, 175:1025--1040, 2021.

\bibitem{Chow2011}
Chi~Wai Chow, Bryan Urquhart, Matthew Lave, Anthony Dominguez, Jan Kleissl,
  Janet Shields, and Byron Washom.
\newblock Intra-hour forecasting with a total sky imager at the uc san diego
  solar energy testbed.
\newblock {\em Solar Energy}, 85(11):2881 -- 2893, 2011.

\bibitem{Redman2018}
Brian~J. Redman, Joseph~A. Shaw, Paul~W. Nugent, R.~Trevor Clark, and Sabino
  Piazzolla.
\newblock Reflective all-sky thermal infrared cloud imager.
\newblock {\em Opt. Express}, 26(9):11276--11283, Apr 2018.

\bibitem{Gohari2014}
M.I. Gohari, B.~Urquhart, H.~Yang, B.~Kurtz, D.~Nguyen, C.W. Chow, M.~Ghonima,
  and J.~Kleissl.
\newblock Comparison of solar power output forecasting performance of the total
  sky imager and the university of california, san diego sky imager.
\newblock {\em Energy Procedia}, 49:2340 -- 2350, 2014.
\newblock Proceedings of the SolarPACES 2013 International Conference.

\bibitem{Marquez2013}
Ricardo Marquez and Carlos~F.M. Coimbra.
\newblock Intra-hour dni forecasting based on cloud tracking image analysis.
\newblock {\em Solar Energy}, 91:327 -- 336, 2013.

\bibitem{Li2012}
Qingyong Li, Weitao Lyu, Jun Yang, and James Wang.
\newblock Thin cloud detection of all-sky images using markov random fields.
\newblock {\em IEEE Geoscience and Remote Sensing Letters}, 9:417--421, 05
  2012.

\bibitem{Fu2013}
Chia-Lin Fu and Hsu-Yung Cheng.
\newblock Predicting solar irradiance with all-sky image features via
  regression.
\newblock {\em Solar Energy}, 97:537 -- 550, 2013.

\bibitem{Liu2015}
S.~{Liu}, L.~{Zhang}, Z.~{Zhang}, C.~{Wang}, and B.~{Xiao}.
\newblock Automatic cloud detection for all-sky images using superpixel
  segmentation.
\newblock {\em IEEE Geoscience and Remote Sensing Letters}, 12(2):354--358, Feb
  2015.

\bibitem{Cheng2017a}
H.-Y. Cheng and C.-L. Lin.
\newblock Cloud detection in all-sky images via multi-scale neighborhood
  features and multiple supervised learning techniques.
\newblock {\em Atmospheric Measurement Techniques}, 10(1):199--208, 2017.

\bibitem{Shi2019}
Chaojun Shi, Yatong Zhou, Bo~Qiu, Jingfei He, Mu~Ding, and Shiya Wei.
\newblock Diurnal and nocturnal cloud segmentation of all-sky imager (asi)
  images using enhancement fully convolutional networks.
\newblock {\em Atmospheric Measurement Techniques}, 12:4713--4724, 09 2019.

\bibitem{CALDAS2019}
M.~Caldas and R.~Alonso-Suárez.
\newblock Very short-term solar irradiance forecast using all-sky imaging and
  real-time irradiance measurements.
\newblock {\em Renewable Energy}, 143:1643 -- 1658, 2019.

\bibitem{HASENBALG2020}
M.~Hasenbalg, P.~Kuhn, S.~Wilbert, B.~Nouri, and A.~Kazantzidis.
\newblock Benchmarking of six cloud segmentation algorithms for ground-based
  all-sky imagers.
\newblock {\em Solar Energy}, 201:596 -- 614, 2020.

\bibitem{MAMMOLI2019}
Andrea Mammoli, Guillermo Terr{\'e}n-Serrano, Anthony Menicucci, Thomas~P
  Caudell, and Manel Mart{\'\i}nez-Ram{\'o}n.
\newblock An experimental method to merge far-field images from multiple
  longwave infrared sensors for short-term solar forecasting.
\newblock {\em Solar Energy}, 187:254--260, 2019.

\bibitem{Nummikoski2013}
Jaro Nummikoski.
\newblock {\em Sky-image based intra-hour solar forecasting using independent
  cloud-motion detection and ray-tracing techniques for cloud shadow and
  irradiance estimation}.
\newblock PhD thesis, University of Texas, San Antonio, 2013.

\bibitem{richardson2017}
Walter Richardson, Hariharan Krishnaswami, Rolando Vega, and Michael Cervantes.
\newblock A low cost, edge computing, all-sky imager for cloud tracking and
  intra-hour irradiance forecasting.
\newblock {\em Sustainability}, 9(4):482, 2017.

\bibitem{nguyen2014}
Dung~Andu Nguyen and Jan Kleissl.
\newblock Stereographic methods for cloud base height determination using two
  sky imagers.
\newblock {\em Solar Energy}, 107:495--509, 2014.

\bibitem{KUHN2018}
P.~Kuhn, M.~Wirtz, N.~Killius, S.~Wilbert, J.L. Bosch, N.~Hanrieder, B.~Nouri,
  J.~Kleissl, L.~Ramirez, M.~Schroedter-Homscheidt, D.~Heinemann,
  A.~Kazantzidis, P.~Blanc, and R.~Pitz-Paal.
\newblock Benchmarking three low-cost, low-maintenance cloud height measurement
  systems and ecmwf cloud heights against a ceilometer.
\newblock {\em Solar Energy}, 168:140--152, 2018.
\newblock Advances in Solar Resource Assessment and Forecasting.

\bibitem{wang2016}
Guang Wang, Ben Kurtz, and Jan Kleissl.
\newblock Cloud base height from sky imager and cloud speed sensor.
\newblock {\em Solar Energy}, 131:208--221, 2016.

\bibitem{wang2019}
Guang~Chao Wang, Bryan Urquhart, and Jan Kleissl.
\newblock Cloud base height estimates from sky imagery and a network of
  pyranometers.
\newblock {\em Solar Energy}, 184:594--609, 2019.

\bibitem{TERREN2020d}
Guillermo Terrén-Serrano, Adnan Bashir, Trilce Estrada, and Manel
  Martínez-Ramón.
\newblock Girasol, a sky imaging and global solar irradiance dataset.
\newblock {\em Data in Brief}, page 106914, 2021.

\bibitem{TERREN2020b}
Guillermo Terrén-Serrano and Manel Martínez-Ramón.
\newblock Multi-layer wind velocity field visualization in infrared images of
  clouds for solar irradiance forecasting.
\newblock {\em Applied Energy}, 288:116656, 2021.

\bibitem{heath1956}
Thomas~Little Heath et~al.
\newblock {\em The thirteen books of Euclid's Elements}.
\newblock Courier Corporation, 1956.

\bibitem{nouri2019}
Bijan Nouri, P~Kuhn, Stefan Wilbert, Natalie Hanrieder, C~Prahl, L~Zarzalejo,
  A~Kazantzidis, Philippe Blanc, and R~Pitz-Paal.
\newblock Cloud height and tracking accuracy of three all sky imager systems
  for individual clouds.
\newblock {\em Solar Energy}, 177:213--228, 2019.

\bibitem{lewis2010}
Peter~M Lewis, Howard Rogers, and Rafe~H Schindler.
\newblock A radiometric all-sky infrared camera (rasicam) for des/ctio.
\newblock In {\em Ground-based and Airborne Instrumentation for Astronomy III},
  volume 7735, page 77353C. International Society for Optics and Photonics,
  2010.

\bibitem{stone1979}
Peter~H Stone and John~H Carlson.
\newblock Atmospheric lapse rate regimes and their parameterization.
\newblock {\em Journal of Atmospheric Sciences}, 36(3):415--423, 1979.

\bibitem{terren2021e}
Guillermo Terrén-Serrano and Manel Martínez-Ramón.
\newblock Processing of global solar irradiance and ground-based infrared sky
  images for solar nowcasting and intra-hour forecasting applications, 2021.

\bibitem{terren2021f}
Guillermo Terrén-Serrano and Manel Martínez-Ramón.
\newblock Review of kernel learning for intra-hour solar forecasting with
  infrared sky images and cloud dynamic feature extraction, 2021.

\end{thebibliography}

\end{document}